\documentclass[10pt]{iopart}

\usepackage{xcolor}
\usepackage{graphicx} 
\usepackage{hyperref}
\begin{document}

\maketitle

\topical[Impact of border defects on the magnetic flux penetration in superconducting films]{Impact of border defects on the magnetic flux penetration in superconducting films }

\author{Alejandro V. Silhanek}
\address{Experimental Physics of Nanostructured Materials, Q-MAT, Department of Physics, Université de Liège, B-4000 Sart Tilman, Belgium}
\ead{asilhanek@uliege.be}

\author{Lu Jiang}
\address{China University of Petroleum, Qingdao, China}
\ead{lujiang@upc.edu.cn}

\author{Cun Xue}
\address{School of Materials Science and Engineering, Northwestern Polytechnical University, Xi’an 710072, China}
\ead{xuecun@nwpu.edu.cn}

\author{Benoît Vanderheyden}
\address{Montefiore Research Unit, Department of Electrical Engineering and Computer Science,
Université de Liège, B-4000 Sart Tilman, Belgium}
\ead{B.Vanderheyden@uliege.be}

\vspace{10pt}
\begin{indented}
\item[]April 2025
\end{indented}

\begin{abstract}
Defects in superconducting systems are ubiquitous and nearly unavoidable. They can vary in nature, geometry, and size, ranging from microscopic-size defects such as dislocations, grain boundaries, twin planes, and oxygen vacancies, to macroscopic-size defects such as segregations, indentations, contamination, cracks, or voids. Irrespective of their type, defects perturb the otherwise laminar flow of electric current, forcing it to deviate from its path. In the best-case scenario,  the associated perturbation can be damped within a distance of the order of the size of the defect if the rigidity of the superconducting state, characterized by the creep exponent $n$, is low. In most cases, however, this perturbation spans macroscopic distances covering the entire superconducting sample and thus dramatically influences the response of the system. In this work, we review the current state of theoretical understanding and experimental evidence on the modification of magnetic flux patterns in superconductors by border defects, including the influence of their geometry, temperature, and applied magnetic field. We scrutinize and contrast the picture emerging from a continuous media standpoint, i.e. ignoring the granularity imposed by the vortex quantization, with that provided by a phenomenological approach dictated by the vortex dynamics. In addition, we discuss the influence of border indentations on the nucleation of thermomagnetic instabilities. Assessing the impact of surface and border defects is of utmost importance for all superconducting technologies, including superconducting resonators, superconducting single-photon detectors, superconducting radio-frequency cavities and accelerators, superconducting cables, superconducting metamaterials, superconducting diodes, and many others.  
\end{abstract}

\ioptwocol

\tableofcontents

\section{Introduction}
\label{Section-1}

\begin{figure}[!ht]
\centering
\includegraphics*[width=\linewidth]{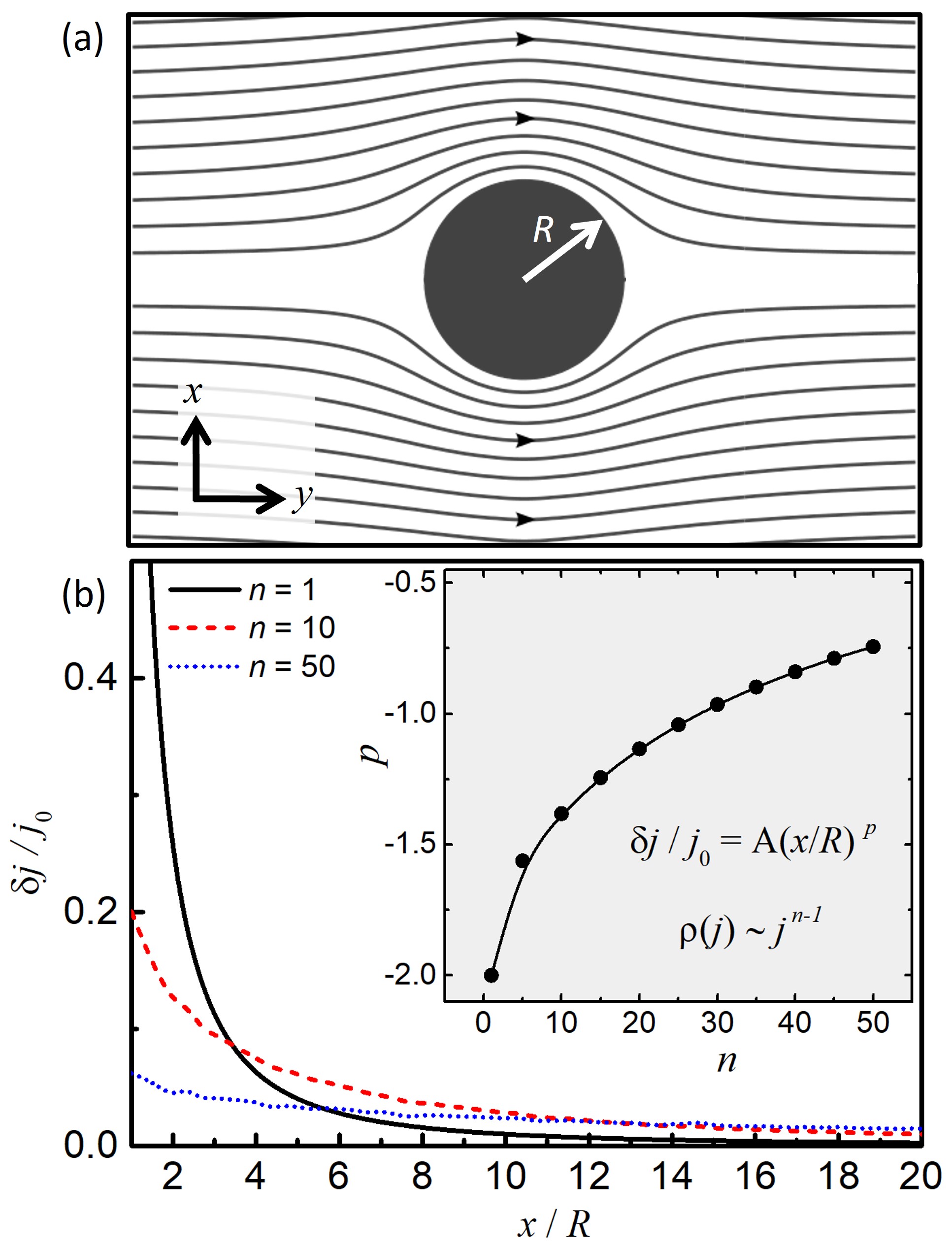}
\caption{(a) current streamlines in a conducting plane with a central hole of radius $R$. (b) The dimensionless perturbation of the norm of the current density, $\delta j / j_0$, as a function of $x/R$, the dimensionless distance from the center of the circular defect. $j_0$ is the value of the uniform current density in a plain sample, without the circular defect. Inset: The decay exponent, $p$, as a function of the $E$-$j$ exponent, $n$ (${\bf E} \sim j^{n-1} \bf{j}$). The exponent $p$ corresponds to the best-fit power-law ($A.x^p$, $A$ being a constant) of the graph $\delta j / j_0$ vs. $x/R$. } \label{Fig1}
\end{figure}

To illustrate how a defect modifies the flow of electric current, let us consider the classical problem of a stationary current flowing in an unbounded normal metallic film. Such a flow is described by a divergenceless current density $\bf{j}$, with $\nabla\cdot\bf{j} = 0$, and an irrotational electric field, $\nabla\times {\bf E} = {\bf 0}$ with ${\bf E} = \rho_m \,\bf{j}$ where $\rho_m$ is the uniform electrical  resistivity of the metal. Thus, the current density vector is similar to the velocity field of a non-viscous incompressible fluid flow. In the  case of an unbounded metallic film, the current is distributed uniformly with a current density $|{\bf j}| = j_0$ and the streamlines are straight and equidistant. The flow is resistive and produces a voltage drop across the film with equipotential surfaces perpendicular to the stream lines. Next, assume that a circular hole is perforated through this metallic sheet, blocking any current into the hole region. Now, the normal component of the current density is vanishing on the hole perimeter and the current must circumvent the hole. The resulting streamlines are distributed symmetrically about the hole, as illustrated in Figure \ref{Fig1}(a). Their density is higher near the hole border, a phenomenon known as \emph{current crowding}~\cite{Hagedorn1963, Clem2011}. One may wonder how far the perturbation in the current density extends into the metallic film. More precisely, we are interested in the distance from the center of the hole and perpendicular to the average current direction that is necessary to recover the unperturbed straight current density. The excess of current density flowing around the hole can be expressed as $(j(x)-j_0)/j_0 = (R/x)^2$, with $R$ the radius of the hole and $x$ the distance from its center perpendicular to the applied current direction. Several interesting remarks can be made here. First, the hole produces a long-range perturbation decaying slowly in space. More specifically, for the deformation of the streamlines to decrease to 1\%, one has to move a distance as far as $10\,R$ from the hole. Second, the result is independent of the conductivity of the sheet, as long as it satisfies Ohm's law. 

A similar perturbation of the current flow can be observed if the film medium is a superconductor with a thickness $d$ much smaller than the London penetration length $\lambda$, and a width $W$ smaller than the Pearl screening length $\Lambda = \lambda^2/d$. In this case, the current density $\bf{j}$ is nearly uniform across the thickness and the sheet current density ${\bf J} = {\bf j} \,d$ satisfies $\nabla\cdot{\bf J} = 0$. Moreover, when $W \ll \Lambda$, it further satisfies $\nabla \times \bf{J} \simeq \bf{0}$~\cite{Clem2011}, so that the resulting current flow also exhibits current crowding at the edge of the hole, with $(j(x)-j_0)/j_0 = (R/x)^2$.

The consequence of current crowding is the suppression of the superconducting state at a reduced critical current, as compared to a similar sample with no hole. The superconducting state was shown to be suppressed through the nucleation of a vortex-antivortex pair when the hole is far from the sample external edges, or through the entrance of vortices via the nearest edge in the other cases. In both cases, the maximal current near the hole is approximately given by the depairing current $j_{dp}$~\cite{Zotova_2014,Vodolazov-2015}. With the law $(j(x)-j_0)/j_0 = (R/x)^2$, current crowding leads to a doubling of the current density at $x = \pm R$, thus high current densities are reached at $x = R$ for a lower net current. It has been observed that the critical current in a strip with a central hole is approximately given by $I_{dp}/2$, where $I_{dp} = W\,d\,j_{dp}$ is the critical current density of the superconducting film without any hole~\cite{Clem2011}. This result clearly points to the effect of a local current crowding on a global property of the system.

Another type of current perturbation can be observed for a hole made
in a type-II superconductor film in the dissipative state. Consider a hole
of radius $R$ in a wide superconducting film ($W \gg R$) with
strong pinning. The hole radius is taken to be much larger than the
typical intervortex separation, and one considers the critical state
formed as a result of the transport of a net stationary current through the
film. A common way to describe the current flow
is by adopting a continuous medium description with
\begin{equation}
  \nabla\cdot{\bf j}=0,\quad\nabla\times{\bf E} ={\bf 0},
  \label{macro-hole-eqs}
\end{equation}
where the electric field $\bf{E}$ is expressed through a constitutive law as
\begin{equation}
  {\bf E} = \rho({\bf j})\,\frac{\bf j}{||{\bf j}||},
  \label{Evsj}
\end{equation}
with the power-law resistivity
\begin{equation}
  \rho({\bf j}) = \rho_0\,\left(\frac{||{\bf j}||}{j_0}\right)^{n-1},
  \label{power-law-rho}
\end{equation}
where ${j_0}$ is some characteristic current density and $n$ is the creep exponent
$n = U_0/k_BT \sim 3 - 50$ where $U_0$ is an energy associated with
depinning, $k_B$ is the Boltzmann constant, and $T$ is the temperature. The electrical resistivity
$\rho({\bf j})$ is expressed over length scales that are both larger than the collective pinning correlation length $L_c \approx (\xi^2 \varepsilon_0^2/\gamma)^{1/3}$~\cite{Blatter1994}  and the intervortex distance $a_0 \approx \sqrt{\Phi_0/B}$, where $\xi$ is the superconducting coherence length, $\varepsilon_0$ is the vortex line energy, $\gamma$ is the disorder parameter, and $\Phi_0$ is the magnetic flux quantum.  Note that the
choice $n = 1$ leads to a medium with an ohmic law, i.e. a metallic
film.

As in the metallic film, the current is assumed to be blocked by the
hole, so that the radial component of ${\bf j}$ vanishes at $r =
R$. As current is injected at a far distance from the hole, the
current density should be flowing uniformly along the $y$-axis for
$|y| \gg R$. As we now show, current crowding is also observed as the
current flows past the hole, but the strong non-linear dependence of
$\rho(\bf{j})$ has a dramatic effect on the spatial extension of the
perturbation.

The stationary problem described by Eqs. (\ref{macro-hole-eqs}),
(\ref{Evsj}), and (\ref{power-law-rho}) is notoriously difficult to
solve~\cite{Gurevich2000}, so that we resort to an auxiliary
problem, solved with a finite element method. The simulation domain is taken as an infinitely long bar extending along the $z$-axis, of square cross section 
$-W/2 \leq x \leq W/2$ and $-W/2 \leq y \leq W/2$, in which a cylindrical hole of
radius $R$ is made parallel to the $z$-axis. The current is driven
by imposing a time-varying magnetic field ${\bf H}$
directed along the z-axis, leading to current densities $\bf{j}$ in
the $(x,y)$ plane. The Maxwell equations
\begin{equation}
\nabla \times {\bf E} = -\frac{\partial {\bf B}}{\partial t},\quad \nabla \times {\bf H} = {\bf 0}, \quad \nabla\cdot{{\bf B}} = 0,
\end{equation}
with ${\bf B} = \mu_0 \, \bf{H}$ are solved while imposing a zero radial
component of ${\bf j}$ around the hole perimeter and a zero current
flow through the external boundaries at $x = \pm W/2$.  These
conditions are imposed by taking a zero magnetic field
${\bf H}$ along the hole perimeter and a field
${\bf H}(\pm W/2,y) = \pm H_a(t)\,\hat{z}$ at the outer lateral
boundaries, forcing a current per unit height $I_w(t) = 2 H_a(t)$ and
an average current density $j_{av} = I_w(t)/W= 2 H_a(t)/W$ through
the system. Moreover, the tangential component of the electric field
is weakly imposed to zero at $y = \pm W/2$, so that the current density
${\bf j}$ is perpendicular to these boundaries.  The average current
density $j_{av}$ is first raised from 0 to $j_0$ as a ramp and is then
kept stationary until transients have decayed. The final current
density distribution is such that $-\partial {\bf B}/\partial t = 0$
and is thus expected to represent the sought stationary current flow.

Simulations are run using the toolkit Life-HTS~\cite{Dular2019}, with
$W/R = 200$, a field applied with a initial rate
$d H_a/dt = 4\pi \times 10^{-9}\times \rho_0 j_0/(\mu_0 R)$, up to a
time $t_1 = 10^{11}/(4 \pi)\times \mu_0 R^2/\rho_0$. It is then left
to relax up to $t_2 = 2 t_1$ to reach the steady state.

Figure \ref{Fig1}(b) depicts the excess current
$\delta j/j_0 = |{\bf j}|/j_0 -1 $ as a function of the dimensionless distance to the
center of the defect, $x/R$, along the $y =
0$ axis.  It can be seen that the higher the index
$n$, the longer the range of the perturbation. In addition, the
current crowding at the border of the hole progressively decreases as
$n$ increases. Note that for all $n$,

\begin{equation}
  \int_R^\infty \frac{\delta j(x)}{j_c}\,dx = R,
\end{equation}

\noindent as the excess current spreading over positions $x > R$ must compensate for the current that would have flown across the radius $R$ in the absence of the hole.

It is possible to approximate each curve in Figure \ref{Fig1}(b) by a decreasing power law, $\delta j/j_0 \approx A.x^p$, with $A$ and $p$ constants. The exponent $p$ characterizes the decrease of the perturbation and exhibits a functional dependence on $n$, as shown in the inset of Figure \ref{Fig1}(b). For $n=1$ (ohmic material), the perturbation decreases quadratically, as anticipated above. In this case, the current density at the border of the defect doubles the applied current density, $j_0$. As $n$ increases, a progressive decrease of the current crowding and a perturbation extending into a larger spatial range are observed. In the limit of $n \rightarrow \infty$, $j(x) = j_0$ so that $\delta j=0$ throughout the sample (no current crowding) and consequently $p=0$. In this limiting case, the norm of the current density is conserved, but the defect still deflects the current, so that the direction of $\bf{j}$ needs to change abruptly along certain lines, known as discontinuity lines, or simply $d$-lines. 

\begin{figure}[!ht]
\centering
\includegraphics*[width=1.0\linewidth]{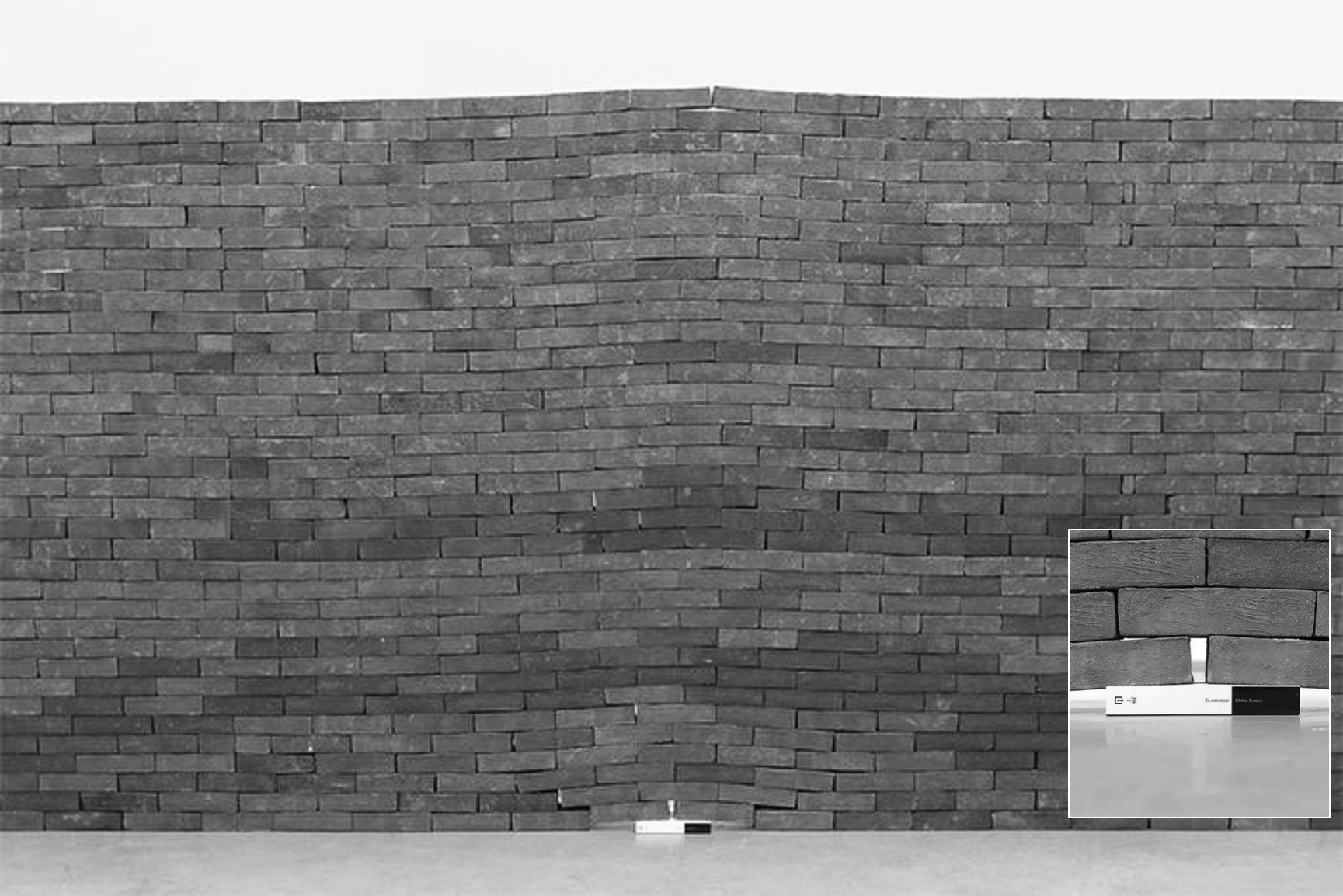}
\caption{A brick wall distorted by a single book by artist Jorge Méndez Blake. At the base of the wall is Frank Kafka’s The Castle (Das Schloss). This art piece illustrates the fact that small perturbations in a rigid medium propagate long distances from the source.} \label{Fig2}
\end{figure}

It may be instructive to draw an analogy between the deformation of current streamlines due to an insulating object and the example depicted in  Figure \ref{Fig2} in which a brick wall with a foreign element (a book) leads to a deformation propagating all the way to the top of the wall following in average a nearly parabolic crack. If the wall were made of a malleable material (or in the extreme case, a fluid) the deformation is expected to be confined to the vicinity of the perturbing element. This analogy makes apparent the fact that the exponent $n$ implicitly bears the concept of rigidity.

The central issue discussed above is a widespread problem that impacts most superconducting devices. Small obstacles can significantly disrupt the two-dimensional flow of current, whether they have microscopic dimensions (the example of a hole in a narrow superconducting strip) or macroscopic dimensions (the example of a hole in a superconductor with strong pinning in a dissipative state). In both examples and in the following discussion, the presence of a defect leads to consequences for the system's current-carrying capacity and macroscopic electrodynamics, including transport and magnetization current. This impact becomes apparent in the modification of global properties such as the critical current or in the form of a long-range decay of electric-field disturbances.

Understanding these phenomena is crucial for comprehending the current-limiting mechanisms in high-temperature superconductors (HTSC) typically containing grain boundaries, second-phase precipitates, microcracks, and areas of local nonstoichiometry. These defects span a broad spectrum of characteristic length scales, ranging from 1 to 1000 nm, which suggest that a multi-physics multi-scale approach is needed to grasp the rich variety of effects emerging from local perturbations. Conventional superconductors and type-I superconductors are not exempted from this problem. From the technological standpoint, border defects play a central role in the performance of superconducting cavities for accelerator applications \cite{Gurevich-2017,Gurevich-2012,Gurevich-2008,Weingarten-2007}, superconducting resonators for quantum computing \cite{Bothner-2011,Bothner-2012a,Bothner-2012b,Chiaro-2016,Kroll-2019,Nulens-2023}, fluxonics devices\cite{Badia-2023,Martinez-2024}, single-photon detectors\cite{sivakov,ZHANG2022}, etc. 

In the present work, we aim to discuss the consequences of border defects on the magnetic flux penetration, focusing on the impact on the electrodynamic properties in thin and thick superconducting films. This review is organized as follows: In Section \ref{Sec2}, we address the problem of single-vortex penetration in a superconducting sample with surface defects. This microscopic picture, as described by London theory and Ginzburg-Landau formalism, is followed in Section \ref{Sec3} by a more macroscopic analysis of the magnetic flux penetration in a superconductor with border defects according to the critical state model. Section \ref{Sec4} reviews and discusses the current understanding of the influence of border imperfections on the triggering of magnetic flux avalanches of thermomagnetic origin. In the concluding section \ref{Sec5}, we outline several pertinent issues that, as per our assessment, still lack resolution and warrant additional research. 

It is essential to emphasize that the literature and references cited in this review should not be regarded as an exhaustive compilation. Given the dynamic and intricate nature of the subject, coupled with the constraints of space in this review, it is inevitable that a considerable number of noteworthy contributions may have been overlooked. Consequently, the references employed here represent a sort of 'working list,' reflecting the publications that the authors of this review have actively engaged with.

\section{Phenomenological picture: Superheating field and vortex penetration in a superconductor} 
\label{Sec2}

\subsection{First vortex penetration without defects}

Individual vortex entry in a type II superconductor becomes thermodynamically favorable when the applied magnetic field $H$ exceeds the lower critical field $H_{c1}$. This critical field is defined as the field above which the Gibbs free energy of a vortex in a superconducting volume becomes negative. However, it has been shown that the Meissner state can remain as a metastable state \cite{Garfunkel-1952,Ginzburg-1958,BL,Fink,Lin-2012,Kramer,Matricon} up to a much higher magnetic field known as supercritical or superheating field  $H_p$. At $H_p$ the Meissner state becomes unstable leading to the formation of vortices and the development of the mixed state. Bean and Livingston \cite{BL} (BL) described, within the London limit, the effect of a flat surface on the vortex penetration into a semi-infinite sample when the vortex axis is parallel to the interface. The London limit is recovered when $\kappa \to \infty$, where $\kappa=\lambda/\xi$ is the Ginzburg-Landau parameter, $\lambda$ is the magnetic penetration depth, and $\xi$ the superconducting coherence length. If the vortex is sufficiently close to the surface, the supercurrent distribution is distorted due to the natural constraint to remain inside the superconducting volume. To satisfy this boundary condition, it is possible to resort to the image method, in which the vortex has an image antivortex equidistant from the interface. As a consequence, an attractive force between the vortex and the surface exists. In addition to this force, the vortex experiences a repulsive Lorentz force at the surface arising from the screening Meissner currents induced by the external magnetic field. The interplay between these competing forces results in the so-called Bean-Livingston surface barrier for vortex entry. According to BL estimation $H_p\approx H_c/\sqrt2$, where $H_c  \approx \kappa H_{c1} /\ln \kappa \gg H_{c1}$ is the thermodynamic critical field. It should be noted that upon entry or exit across the surface barrier, vortices moving in or out of the sample dissipate energy. The surface barrier may thus contribute a sizeable part of the AC losses, as shown in Ref.~\cite{Clem1979}.

As recognized by Bean and Livingston in their original article, the used London formalism ignores the decrease of the density of states with increasing current. This effect is properly described within the Ginzburg-Landau formalism, which couples the Cooper pair density with the pair velocity. It has been suggested within the Ginzburg-Landau and Eilenberger equations \cite{Fink,Lin-2012,Kramer,Aslamazov,Vodolazov99} that the onset of instability of the Meissner state is achieved when the screening current density at an ideal defect-free surface approaches the depairing current density $j_{dp}$. This prediction seems to be in agreement with experimental measurements of the first vortex penetration in Pb thin films via scanning Hall probe microscopy \cite{Gutierrez}.

It should be pointed out that the Bean-Livingston approximation considers an already nucleated vortex without analyzing the vortex nucleation process. This process is discussed by Galaiko \cite{Galaiko} who analyzed the stability of the superconducting state with respect to fluctuations of the order parameter and calculated the critical dimensions and the shape of the vortex nucleus that is produced near the surface of the superconductor, as well as the energy barrier associated to the nucleus. This author suggested that the key parameter delaying the vortex penetration is the elastic energy of the vortex line, rather than the vortex image. 

\begin{figure}[!ht]
\centering
\includegraphics*[width=1.0\linewidth]{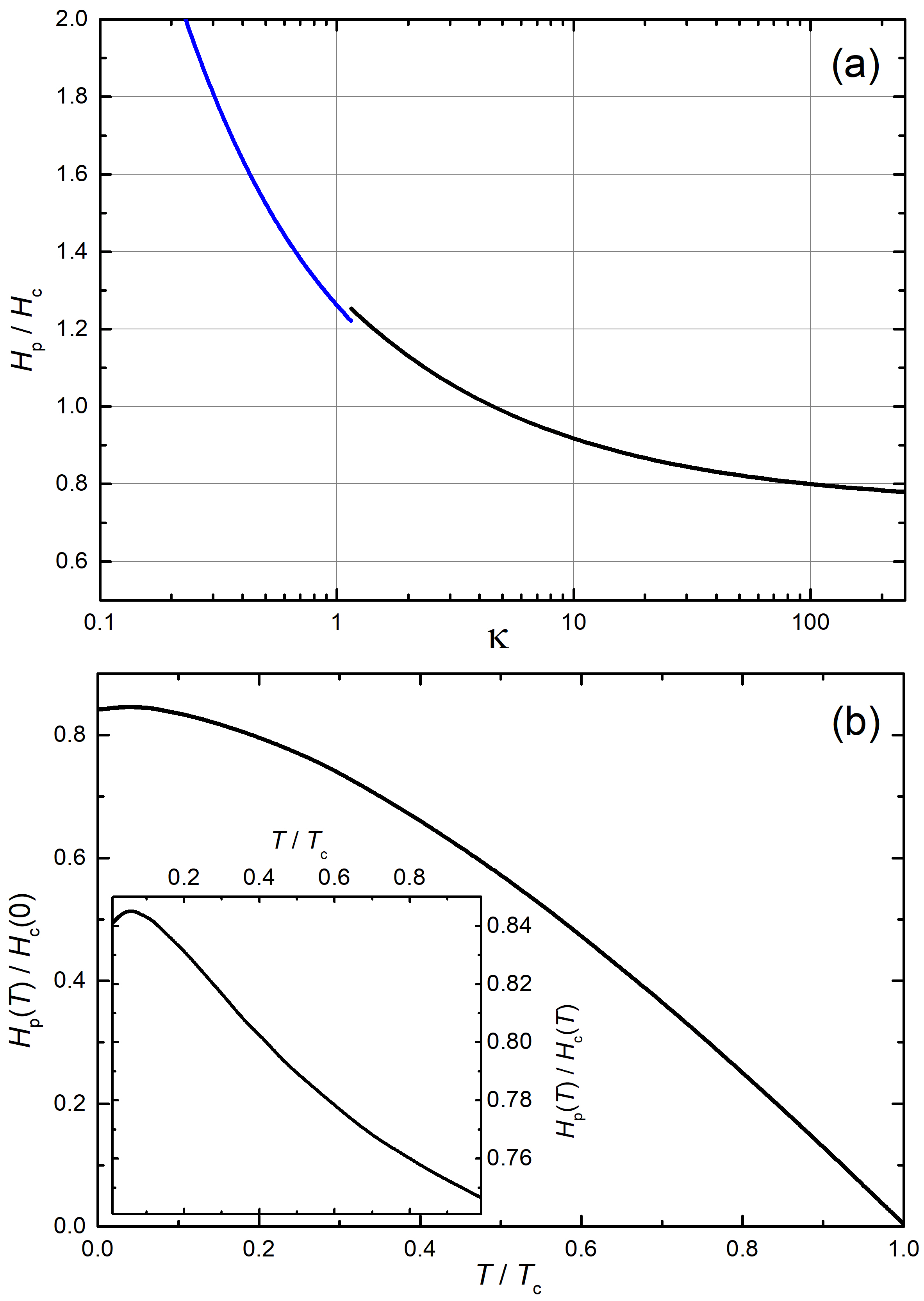}
\caption{(a) Superheating field $H_p$ as a function of the Ginzburg-Landau parameter $\kappa$ using analytical approximations corresponding to the regimes of one-dimensional (long wavelength of the instability of the Meissner state) critical perturbations ($\kappa < 1.1495$, blue curve) and two-dimensional (short wavelength) critical perturbations ($\kappa > 1.1495$, black curve). Adapted from Ref.\cite{Transtrum}. (b) Superheating field as a function of the reduced temperature $T/T_c$ normalized by the critical field at zero temperature (main panel) and by $H_c(T)$ (inset). Adapted from Ref.\cite{Catelani-2008}} \label{Fig3}
\end{figure}

Calculations of $H_p$ based on Ginzburg–Landau theory \cite{Kramer,chapman} have shown that the penetration field lies slightly above the limit imposed by the BL model, with $H_p \approx 0.745 H_c$ (only 5\% above the BL prediction) if $\kappa \rightarrow \infty$ and grows as $\kappa$ decreases \cite{deGennes-1965,Kramer,Fink} with $H_p \approx H_c$ when $\kappa=4.25$ \cite{Fink}. Quantitative experimental data seem to agree with these predictions \cite{Boato,DeBlois,Yogi}. The theoretical $H_p (\kappa)$ dependence extracted from Ref.\cite{Transtrum} is shown in Figure \ref{Fig3}(a). In Ref.\cite{Transtrum}, it is shown that the breakdown of the Meissner state at $H_p$ takes place via a periodic perturbation of the order parameter with a wavelength $(\xi^3 \lambda)^{1/4}$ along the surface. A similar decrease of $H_p(\kappa)$ was already captured by Matricon and Saint-James as early as 1967 \cite{Matricon} as well as by Fink and Presson in 1969 \cite{Fink}.

Since GL theory is applicable close to the superconducting critical temperature (i.e., if $T_c-T \ll T_c$), a different approach is needed to analyse the stability of the Meissner state at low temperatures. This has been done in Ref.\cite{Catelani-2008} by implementing Eilenberger’s semiclassical approximation for a superconducting system in the clean limit. These authors have calculated $H_p$ as a function of temperature for $\kappa \gg 1$ (see Figure \ref{Fig3}(b)) and shown that the thermodynamic critical fields is a nonmonotonic function of temperature exhibiting a maximum at $T \approx 0.06\,T_c$ (thus clearly illustrating the risk of extrapolating the GL solution to zero temperature). These results were extended to the case of superconductors with magnetic and nonmagnetic impurities by Lin and Gurevich \cite{Lin-2012}, who showed that nonmagnetic impurities suppress the weak maximum in $H_p$ at low temperatures. These authors have also shown that nonmagnetic impurities weakly affect $H_p$ even in the dirty limit, while magnetic impurities suppress $H_p$ and $T_c$.

\begin{figure}[!ht]
\centering
\includegraphics*[width=1.0\linewidth]{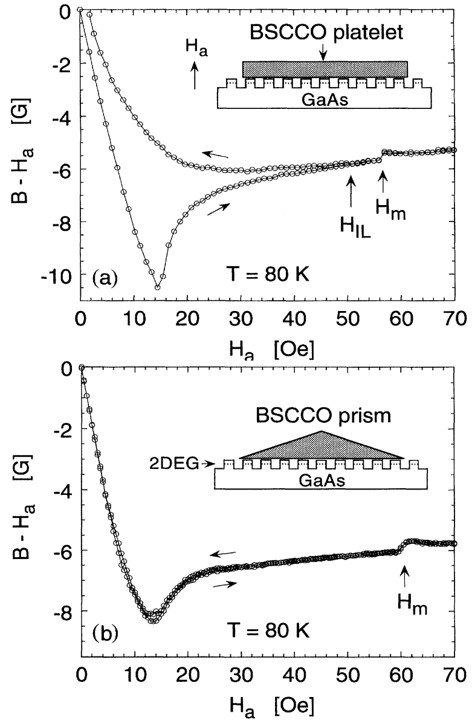}
\caption{Local magnetization loops $B-H_a$ as a function of applied magnetic field $H_a$ in BSCCO crystals of platelet (a) and prism (b) shapes at $T = 80$ K. The platelet crystal shows hysteretic magnetization below the irreversibility field $H_{IL}$. In the prism sample, the geometrical barrier is strongly suppressed, and a fully reversible magnetization is obtained at temperatures above 76 K. The vortex melting transition $H_m$ is observed as a sharp thermodynamic step in the local magnetization. A cross-section of the experimental setup is shown schematically in the insets (not to scale). The two-dimensional electron gas active layer of the sensors resides about 0.1 µm below the surface. The BSCCO crystals are in contact with the GaAs surface, and the local vertical component of the magnetic field $B$ is measured directly. The external field $H_a$, is applied parallel to the crystalline $c$-axis. Reproduced from Ref.\cite{Majer1995}.} \label{Geometrical-Barrier}
\end{figure}

The sample geometry may also play an important role in the vortex penetration, since it influences the current distribution within and at the edges of the sample. The superheating field in geometries other than semi-infinite superconducting half-space, such as finite cylinders \cite{Fink-1968} and slabs \cite{deGennes-1965,Fink-1967}, has also been investigated theoretically. For a cylinder of radius $R \geq 20\lambda$, the solution is similar to that of a semi-infinite superconducting half-space, whereas for smaller radius, the superheating field exceeds that expected for a planar surface. In films with a large demagnetization factor in a perpendicular magnetic field, the strong curvature of the field lines at the edges results in a pronounced inhomogeneous current distribution. When the sample is thicker than the penetration depth and it has a constant thickness, a competition between the line tension of a vortex cutting through the upper and lower ridges at the sample edge and the Lorentz force induced by the Meissner currents results in a penetration barrier known as {\it geometrical barrier} \cite{Zeldov94,Benkraouda}. For sample thicknesses smaller than the penetration depth, this effect is unimportant and BL-type barriers set the penetration field \cite{Zeldov-2013}. Superconductors of ellipsoidal shape do not exhibit this type of surface barrier, as the line tension force is exactly balanced by the Lorentz force. Such samples have thus been used to determine the lower critical field $H_{c1}$ \cite{Liang94}. A theoretical approach to evaluate the interplay between line tension and Lorentz forces for an arbitrary shape of the sample was reported in~\cite{LabuschDoyle1997}. It should be emphasized that the geometrical barrier is strongly suppressed if the sample thickness increases over its entire half-width.  Thus, polishing the sample in the shape of a prism substantially decreases the geometrical barrier~\cite{Majer1995}, as shown in Fig. \ref{Geometrical-Barrier}. For other types of geometries, the shape of the sample generally results in a more prominent penetration barrier. It has been shown that geometrical edge barriers can play a dominant role in the increase of the critical current density of YBa$_2$Cu$_3$O$_{7-\delta}$ (YBCO) thin films as thickness decreases \cite{jones-2010,rouco-2019}. Bi$_2$Sr$_2$CaCu$_2$0$_{8+\delta}$ (BSCCO) single crystal platelet with irradiation-enhanced pinning in its edge zone (i.e. a framelike zone of enhanced bulk pinning), was used to simulate a geometrical barrier for flux penetration \cite{Schuster-PRL-1994}. The obtained flux-density profiles in increasing applied field agree with those calculated by assuming $H_{c1}=0$ and a nonlinear current-voltage law with two different critical current densities in the irradiated and unirradiated zones, and zero electrical resistance at positions where $B = 0$. The magneto-optical images of the magnetic field landscape obtained in Ref.\cite{Schuster-PRL-1994} show that the magnetic flux, after crossing the irradiated frame, piles up and forms a flux dome at the center of the sample. This situation is analogous to the penetration of flux bundles of about 60$\Phi_0$ over an edge barrier observed in type-I Pb superconductors \cite{Huebener1972}.

\begin{figure*}[!ht]
\centering
\includegraphics*[width=1.0\linewidth]{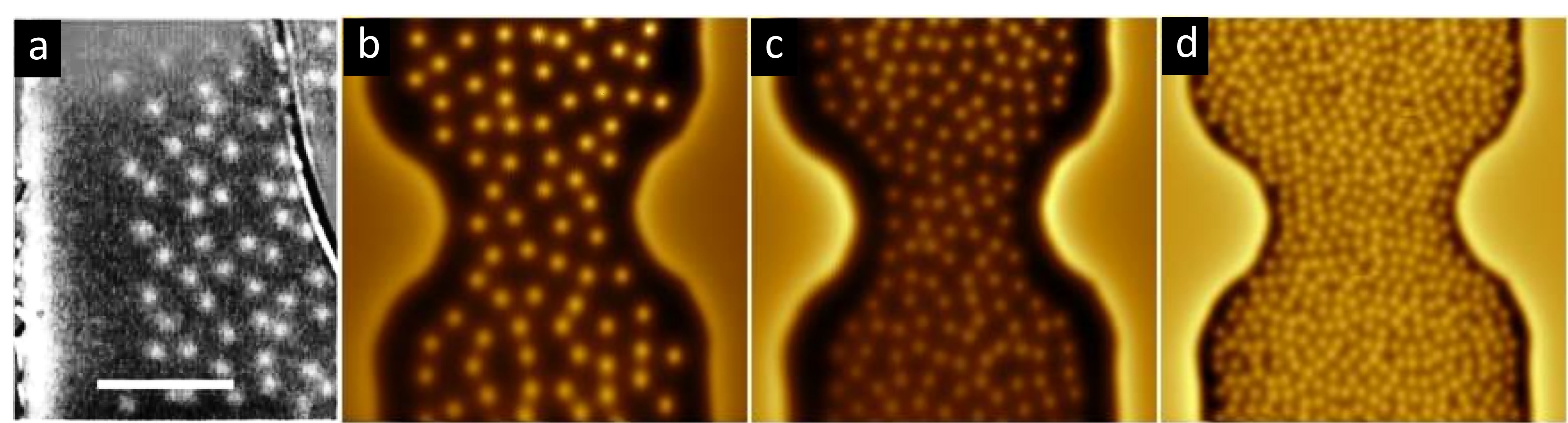}
\caption{ (a) Magneto-optical image showing the vortex penetration in zero-field cooling condition in a 100 µm thick NbSe$_2$ single crystal, at an applied magnetic field of 0.2 mT. The vortex-free region is seen between the edge (bright vertical band on the left) and the vortex-filled region. The scale bar corresponds to 10 µm. Reroduced from Ref.\cite{Olsen}. Scanning SQUID microscopy covering a surface of 12 µm $\times$ 12 µm obtained in a Pb film with a 5.7 µm-wide central constriction at 4.2 K and after field-cooling in fields of (b) 2.7 mT, (c) 5.4 mT, and (d) 12 mT. The vortex-free region is seen to shrink as the magnetic field increases. Reproduced from Ref.\cite{Embon}.}\label{vortex-free}
\end{figure*}

In high-$T_c$ superconductors, vortices can overcome the BL barrier by thermal activation \cite{Buzdin-Feinberg}, however, the geometric barrier cannot be overcome by thermal activation and is less sensitive to surface roughness.  This subject is amply discussed by Brandt \cite{Brandt-review} where at least seven different types of surface barriers are identified. It is worth noting that the presence of BL or geometrical barriers leads to vortex-free regions along the edges of the sample \cite{Ternovskii,Clem1974,Burlachkov-1993,GorbachevSavelev1996,Kuznetsov99,Elistratov2002}. This vortex-free region has been observed with single vortex resolution in NbSe$_2$ single crystals \cite{Olsen}, and Pb constrictions \cite{Embon} as illustrated in Figure \ref{vortex-free}. Since a screening current flows near the surface of the sample, and the number of nearest neighbors for each vortex line located near the surface is different than for those in the interior, it must be expected that the lattice of vortex lines will be deformed near the surface of the sample \cite{Ternovskii} even in absence of pinning. 

The problem of vortex entry barrier needs to be revisited when dealing with thin films ($d<\lambda$) in a perpendicular field. The reason is that vortices in films interact mostly via the stray fields in the surrounding space, giving rise to longer-range interactions \cite{Pearl}. In this case the screening of the in-plane supercurrents is governed by the effective penetration depth $\Lambda = \lambda^2/d>\lambda$ \cite{Carneiro,Brandt}. In addition, the method of images as used by Bean-Livingston cannot be applied in thin film geometry, and the attraction of a vortex to the film edges develops on a scale noticeably larger than $\Lambda$ \cite{Kogan94}. This problem and the competition between BL barriers and geometrical barriers is discussed in Ref. \cite{Zeldov-2013,Mikitik-2021} for the case of a pinning-free superconductor with rectangular cross-section and in the limit where the width and the thickness of the strip are larger than the London penetration depth. The key parameter describing the interplay between BL and geometrical barrier is,

\begin{equation}
  p \equiv \frac{\kappa}{\ln \kappa} \left(  \frac{\lambda}{d} \right)^{1/3} ,
\end{equation}

\noindent which is to be compared with $p_c \approx 0.52$ for thin isotropic strips (for anisotropic slabs, $p_c$ depends weakly on the aspect ratio $d/W$, with $W$ the width of the slab) \cite{Mikitik-2024}. If $p>p_c$, then the Bean-Livingston barrier in the corners of the strip determines the critical current at zero applied magnetic field. Otherwise, for $p<p_c$, the critical current is determined by the geometrical
barrier. Since $p$ increases with decreasing thickness, it is expected that the Bean-Livingston barrier dominates as the thickness decreases. For the sake of illustration on the interplay between the different types of barriers, let us have a look to the case of Nb thin film for which
$d/W \ll 1$, taking the parameters $\lambda \approx 92$ nm, $\xi \approx 12$ nm \cite{Motta2014}, we estimate $d_c \approx 35$ µm,
suggesting that the flux penetration in Nb thin films is dominated by Bean-Livingston barriers.

Interestingly, the extreme limit $d \ll \lambda$ was treated
analytically in Refs.~\cite{MaximovElistratov1995}, while taking into
account bulk pinning. It was further extended to situations with
transport current in Ref. \cite{MaximovEPL1995}. Even though the
Bean-Livingston barrier is dominant in this limit, the vortex
penetration processes resemble closely those observed for a purely
geometrical barrier: for a zero transport current and weak pinning,
the magnetic flux accumulates in the central region of the film
whereas Meissner currents arise in the peripheral regions. The
  surface barrier in the regime with $d \ll \lambda$ has been named the {\it extended Bean-Livingston barrier} or {\it edge barrier}.  Further analysis of the relative importance of the edge barrier and the bulk pinning on the magnetization and the critical current was studied in Refs.~\cite{Elistratov2000,Elistratov2002}. The case of a thin film of width $W$ comparable to the effective penetration depth
  $\Lambda = \lambda^2/d$ is treated for a weakly pinned superconductor in Ref.~\cite{Plourde2001}.

In the case of a thin film with an applied field perpendicular to the plane of the film and assuming $\Lambda \gg \lambda$, so as to neglect the magnetic field created by the currents, numerical simulation of the time-dependent Ginzburg-Landau equation by Aranson \etal \cite{Aranson-95,Aranson-96} showed that the flux penetration into a thin superconducting film occurs via the dynamic suppression of the order parameter on the macroscopic scale, producing nucleation of extended droplets of the normal phase in the superconducting sample. The droplets contain multiple topological charges and are therefore unstable with respect to splitting into singly charged vortices, particularly when pinning is present. The time scale of formation and splitting of the droplets is on the order of ns. 

The onset of flux penetration (i.e. after ZFC) into a thin superconducting film strip of width $W$ has been investigated by Kuznetsov \etal \cite{Kuznetsov99} in the limit $W \gg \Lambda$ and for thick films dominated by geometric barriers \cite{Kuznetsov97}. These authors also discuss the influence of border roughness and polycristalline samples and show that the effect of pinning and roughness is to lower the penetration field. Further theoretical investigations have shown that (i) surface barriers are also present in type I superconductors \cite{Kuznetsov-1998,Castro-1999}, (ii) a normal (rather than insulating) boundary condition reduces the surface barrier and makes it nearly independent of $\kappa$ \cite{Hernandez}, and (iii) $H_p$ increases as the dimension of the sample decreases and is largely underestimated by London model calculations \cite{Hernandez,Berdiyorov}. The case of vortex trapping and expulsion (i.e. after FC) has been largely addressed in the literature \cite{Likharev-1972,Maksimova-1998,Clem-1998,Field-2004,Bronson-2006,Kuit,Kramer-2010,Ge-2023}. Both experiments and theory found that the onset field for vortex trapping when $W \ll \Lambda$ varied roughly as 1/$W^2$.
 
It is worth mentioning that the estimation $H_p \sim H_c$ remains valid for anisotropic superconductors. However, in the extreme anisotropic case in which nearly decoupled superconducting $x$-$y$ planes can be considered in the quasi-2D regime, the transverse coherence length $\xi_z$ becomes smaller than the interlayer distance $d$, and the field orientation is parallel to the layers, then $H_p$ can be substantially lower than $H_c$ \cite{Buzdin-Feinberg}. It has been experimentally demonstrated that edge barriers play a prominent role in magnetization processes \cite{Kopylov} and transport properties \cite{zeldov} of single crystals of HTSC superconductors. In general, it is not straightforward to separate effects coming from the surface (BL barriers or GB) from those originating from bulk pinning. Some successful approaches to eliminate the geometrical barrier include polishing a single crystal into a prism shape \cite{Majer1995}, tilting the magnetic field in anisotropic superconductors \cite{segev}, or using the field-focusing effect in a double-strip with a narrow gap geometry \cite{willa}.

It has been recently predicted theoretically that dirty superconductors with weak volume pinning and strong edge barrier for vortex entry should promote ultra-fast vortex dynamics \cite{Vodolazov-2019}. This is particularly relevant for the enhancement of superconducting single-photon detectors, which should be biased by close-to-depairing critical currents. These conditions have been met experimentally in Pb \cite{Embon} and Nb-C \cite{Dobrovolskiy-2020} bridges. The former in a regime of nonuniform current distribution ($\Lambda \ll W$) and the latter for uniform current distribution ($\Lambda \gg W$).

In multicomponent superconductors (multiband, type 1.5, etc.) vortices with a winding in the phase of only one of the components can be stable topological solitons that carry a fraction of the flux quantum. These objects were shown to be stable near the sample boundaries in the two-component London model. Therefore, the conventional Bean-Livingston picture of magnetic flux entry needs to be revisited, as discussed in Ref.\cite{Maiani}. Another interesting effect predicted for d-wave superconductors is the formation of localized Andreev bound states, situated at a distance $\sim \xi$ from the edges of the sample \cite{Fogelstrom,Hu}. These zero-energy excitations develop when the sample edge is oriented along the nodal direction of the d-wave superconducting order parameter and generate an anomalous Meissner current running opposite to the supercurrents \cite{Iniotakis-2005}. This leads to a decrease in the net surface current for samples with edges parallel to the nodal direction that, in turn, enhances the Bean-Livingston barrier since the force pushing the vortex inside the sample decreases whereas the attraction towards the edge remains unaltered. Therefore, a dependence of $H_p$ on the crystal orientation of the sample edge is expected \cite{Iniotakis-2008}. Some experimental evidence supporting this prediction has been reported in Ref. \cite{Doltz}.

\subsection{First vortex penetration with border defects (London theory)}

Although the Meissner phase can, in principle, survive as a metastable state up to a penetration field $H_p \sim H_c$, experimentally this upper limit is hardly reached. The reason behind this observation, as for any thermodynamic metastability, lies in the presence of nucleation points where the local value of the Gibbs free energy is reduced, thus favoring the development of the most stable phase. In a superconductor, this corresponds to small surface imperfections, oxidation, grain boundaries, and other defects where the BL barrier is partially suppressed and so is $H_p$ \cite{Burlachkov92}. Since the vortex self-energy and the screening currents primarily change in a narrow region of width $\sim$ max\{$\lambda,\Lambda$\} from the edge, it is natural to think that local damage in this region can severely affect the strength of the penetration barrier \cite{Kuznetsov99}.  

\begin{figure}[!ht]
\centering
\includegraphics*[width=1.0\linewidth]{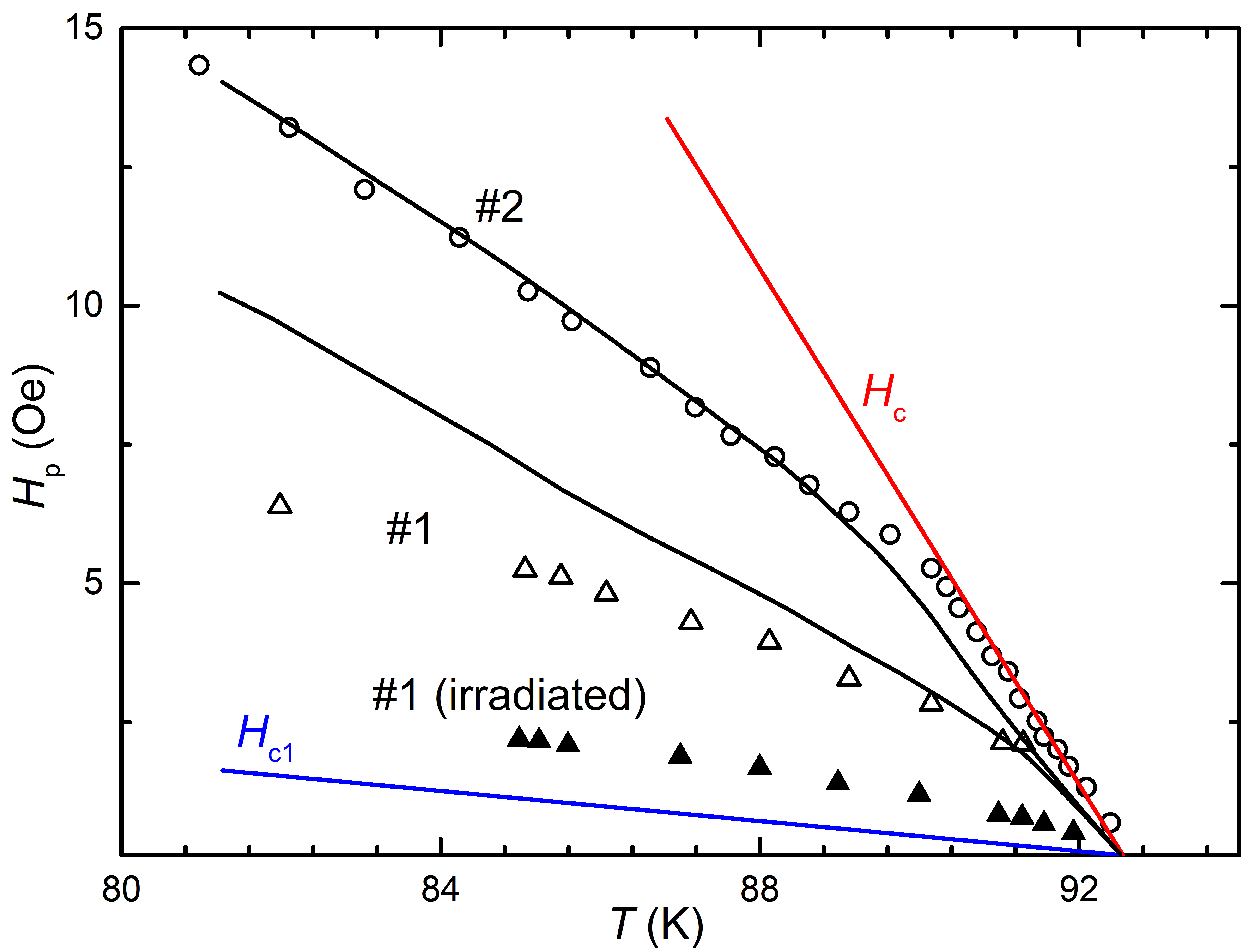}
\caption{Temperature dependence of the first vortex penetration field $H_p$ for two untwinned samples (\#1 and \#2) in the as-grown state (open symbols) and after electron irradiation (solid symbols). Note that the scale of fields depends on the demagnetization factor. The solid curves represent theoretical fits to $H_p$ in terms of BL barriers, whereas $H_c$ and $H_{c1}$ have been determined using $\lambda_0$ = 170 nm and $\xi_0$ = 2 nm. Adapted from Ref. \cite{Burlachkov91}} \label{Burlachkov}
\end{figure}

In 1991 Burlachkov \etal  \cite{Burlachkov91} investigated the penetration field in the HTSC material YBCO motivated by the large value of $\kappa \sim 100 $ thus permitting a sizable separation between the lower critical field $H_{c1}$ and the maximum penetration field $H_p=\kappa H_{c1}/\ln \kappa \approx 22 H_{c1}$. Miniature Hall probes were used to determine the onset of vortex penetration into the sample. In Fig. \ref{Burlachkov} the obtained $H_p (T)$ is plotted for two untwinned single crystals. At high temperatures, $H_p(T)$ shows a linear increase as $T$ decreases which turns to a nonlinear dependence at lower $T$. This change of regime has been attributed to the presence of microscopic surface defects of size $a \sim \xi(0)$, where $\xi(T)$ is the temperature-dependent coherence length. At high temperatures, both $\lambda(T)$ and  $\xi(T)$ are much larger than $a$ and the defects do not influence the surface barrier, hence $H_p \sim H_c$. At lower temperatures, the vortex size becomes comparable to the defect size, and the penetration field drops to some intermediate value. In addition, the authors show that a low dose of 2.5 MeV electron irradiation significantly reduces $H_p(T)$. This result has been interpreted in terms of defects created in the bulk during irradiation, which then migrate to the surface during room-temperature annealing. The effect of irradiation has been further investigated in Ref.\cite{Konczykowski91,Chikumoto91}.

\begin{figure}[!ht]
\centering
\includegraphics*[width=1.0\linewidth]{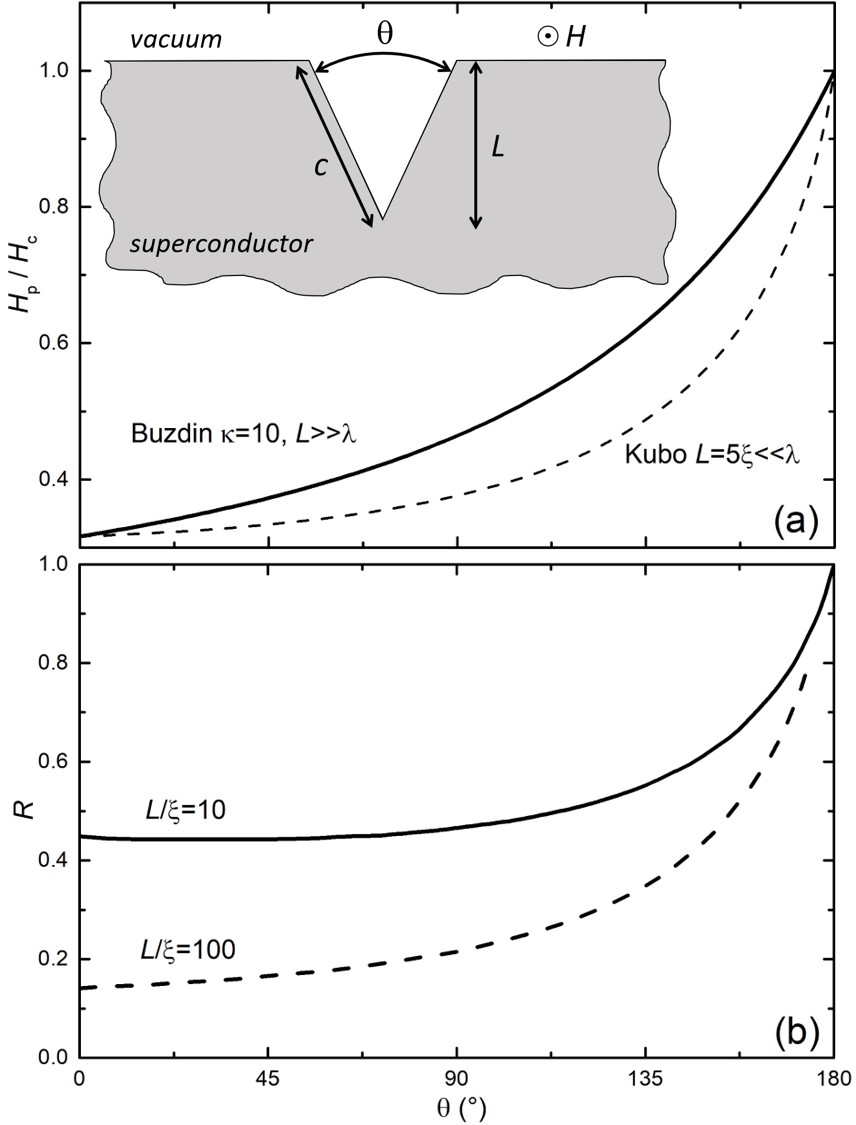}
\caption{(a) Suppression factor of the penetration field as the angle $\theta$ of the border indentation according to Buzdin and Daumens \cite{Buzdin98} and Kubo's \cite{Kubo} expressions for the particular cases of $\kappa=10$ and depth of the indentation $L=5\xi$. Inset: schematic drawing of a triangular border indentation and the geometrical parameters involved. (b) Reduction factor of the critical current $R=j_c/j_0$ as a function of the vertex angle $\theta$ for two different values of the ratio $L/\xi$. Adapted from Ref.\cite{Clem2011}.} \label{Buzdin}
\end{figure}

The influence of small size ($< \lambda$) surface defects on the Bean-Livingston barrier for a single Abrikosov vortex has been calculated within the London limit by Bass \etal \cite{Bass96}. It is shown that even a rather smooth surface roughness results in an essential decrease of the BL surface barrier. For example, a root-mean-square surface roughness of $0.4\lambda$ decreases by 10\% the energy barrier from its value for the ideal surface. The case of large size defects ($> \lambda$) was addressed theoretically in Refs. \cite{Buzdin98,Aladyshkin2001}. The interaction of a vortex with a border defect was treated by Buzdin and Daumens \cite{Buzdin98} within the two-dimensional London formalism and hence more suitable for high-$\kappa$ superconductors such as HTSC. They calculated the first vortex penetration field $H_p$ near a triangular border indentation of internal vertex angle $\theta$ (see Fig.\ref{Buzdin}). The presence of the defect increases the Lorentz force towards the center of the sample due to current crowding and, at the same time, diminishes the attractive force caused by the vortex current deformation. Both effects tend to decrease the penetration field to $H_p \approx \gamma H_c /\kappa^{\frac{\pi-\theta}{2\pi-\theta}}\ll H_c$ with $\gamma \sim 1$. This result has later on been refined by Aladyshkin \etal \cite{Aladyshkin2001} who found $\gamma = 1$ for an ideal plane surface ($\theta=\pi$) and  $\gamma = \sqrt{\pi}$ for a thin crack (i.e. $\theta \ll 1$). More recently, an analytical expression for $H_p/H_c$ for the case of a triangular groove of length $\xi \ll L \ll \lambda$ has been derived by Kubo \cite{Kubo}. In Figure \ref{Buzdin} the reduction of the penetration field as the angle $\theta$ of the border indentation decreases, is plotted using Buzdin and Daumens's expression \cite{Buzdin98} for the case $\kappa=10$, $L\gg \lambda$, and Kubo's expression \cite{Kubo} for an indentation depth $L=5\xi$. 

A related problem is that of the calculation of the current density $j_{0}$ (instead of the magnetic field) needed to reduce the Gibbs-free-energy barrier to zero allowing vortex penetration into the sample. Clem and Berggren \cite{Clem2011} have recently tackled this question within the London model, for the case of a superconducting strip of thickness $d\ll \lambda$ with a border indentation, and where its width is much smaller than the Pearl length $\Lambda= \lambda^2/d$. For a straight strip without defects $j_{0}=\phi_0/e\pi\mu_0\xi d \Lambda \approx H_c/\lambda$ which corresponds to the depairing current $j_{dp}$. In presence of a border defect, due to the current crowding the critical current density $j_c$ is reduced by a factor $R<1$ such that $j_c=Rj_0$. This factor as function of $\theta$ for a triangular indentation is shown in the lower panel of Fig.\ref{Buzdin} for three different values of the ratio $\xi/L$ and follows the functional form,

 \begin{equation} 
R=\frac{2\pi-\theta}{\pi}\left( \frac{\xi}{2gc} \right)^{\frac{\pi-\theta}{2\pi-\theta}},
 \end{equation}
 
\noindent where $g(\theta)=1$ at $\theta=0$ and $\pi$, and rises smoothly to a maximum of $g=1.180$ at $\theta=\pi/2$ and $c=L/\cos(\theta/2)$.

The authors of Ref.\cite{Clem2011} also calculated the reduction of the critical current when the notch has a semicircular form of radius $a$. When $\xi \ll a$, the critical-current reduction factor is $R = 1/2$, which arises from current crowding at the top of the notch where the sheet-current density is a factor of two larger than far away from the notch (see Section \ref{Section-1}). When the linear dimensions of the edge defect are much smaller than $\xi$, the suppression of the critical current is negligible and $R \approx 1$, as expected.

\subsection{First vortex penetration with border defects (Ginzburg-Landau theory)}

The calculations of the vortex penetration in a notched superconductor discussed so far have made use of the London model which implicitly assumes that the superconducting order parameter is not suppressed by the current density. The accuracy of the presented results could therefore  be improved by using the Ginzburg-Landau formalism applicable at temperatures close to the superconducting transition. 

\begin{figure}[!ht]
\centering
\includegraphics*[width=\linewidth]{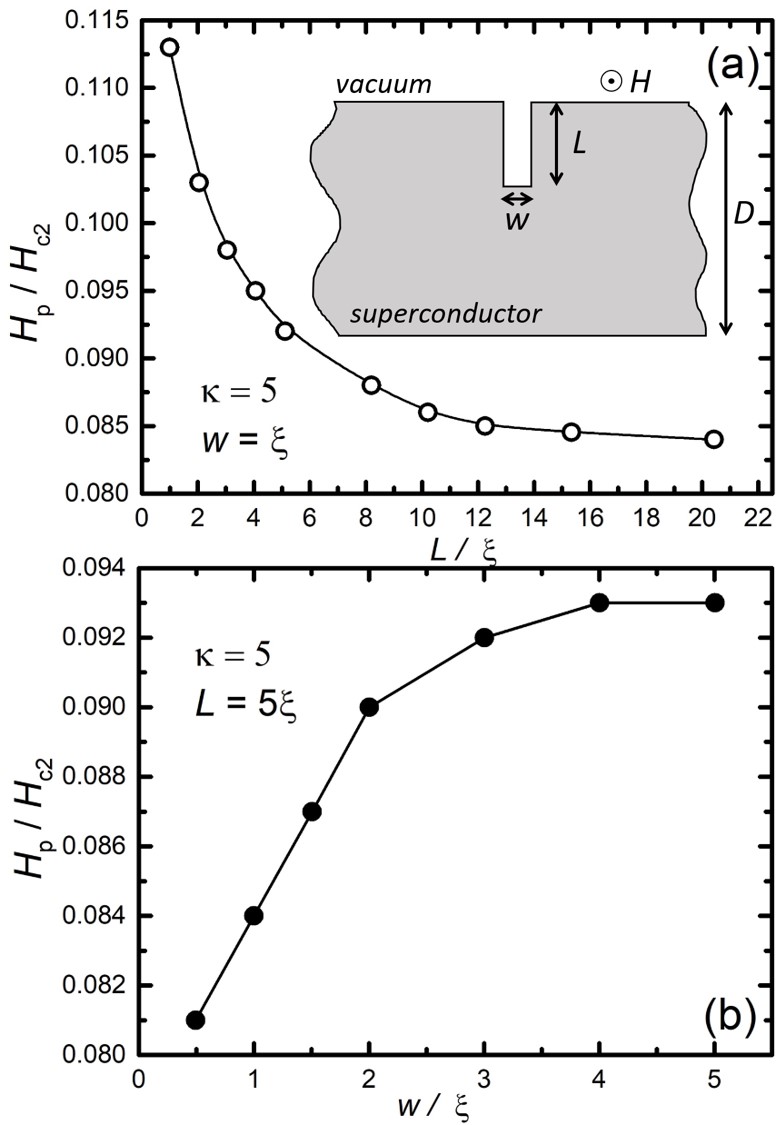}
\caption{Inset in panel (a) shows the geometry of the considered system. The sample extends in horizontal direction and along the magnetic field direction. (a) Dependence of the entry field $H_{p}$ on a defect length for a defect width $w=\xi$ for type-II bulk superconductor ($\kappa =$5). (b) Dependence of field entry field $H_{p}$ on a defect width for a defect length $L=5\xi$. Adapted from Ref. \cite{Vodolazov00}} \label{Vodolazov}
\end{figure}

Numerical and analytical solutions of the GL equations applied to describe the vortex entry conditions in a bulk superconductor with a surface defect have been found by Vodolazov \etal \cite{Vodolazov00,Vodolazov03}. Ref. \cite{Vodolazov00} study the effect of
rectangular-shaped surface defects on the vortex entry in {\it bulk} type-II superconductors. The geometry of the considered system is shown in the inset of Fig.\ref{Vodolazov}(a). Panel (a) shows the dependence of the entry field $H_{p}$ as a function of defect length $L$ and for a defect width $w=\xi$ for a type-II superconductor ($\kappa = 5$). The panel (b) shows the dependence of $H_{p}$ on defect width $w$, for a defect length $L=5\xi$. It is shown that the field $H_{p}$ decreases monotonically with a growing length and a decreasing width of the rectangular notch. It was also found that the penetration field $H_{p}$ reaches a minimum possible value (i.e. the most detrimental defect) for defects with a length $L \geq 2\lambda$ and a width $w \approx \xi$. In this particular limit, the author provides interpolation expressions for $H_{p}$, faithfully reproducing the numerical results with a high degree of accuracy (error smaller than 2\%),

\begin{eqnarray}
\frac{H_{p}}{H_c} \approx \frac{0.96}{\kappa^{1/3}} ~~~~~~  1/\sqrt{2} \leq \kappa \leq 5, \label{eq1}\\
\frac{H_{p}}{H_c} \approx \frac{1.03\sqrt{2}}{\kappa^{1/2}}(1-0.63/\kappa)  ~~~~~~  5 \leq \kappa \leq 20. \label{eq2} 
\end{eqnarray}

In Ref.~\cite{Vodolazov03} the authors consider two types of square shape defects in a finite size sample, (i) inclusions on the superconductor surface having a lower $T_c$ than the bulk, and (ii) roughness produced by geometrical irregularities. For the former, inclusion sizes ranging from $\xi \times \xi$ up to  $8\xi \times 8\xi$ were analyzed and an entry field of about $0.3H_{c2} \sim 0.85H_{c} $ were reported for all sizes and different variations of the local $T_c$ and $\kappa=2$. Interestingly, it is found that inclusions of another phase give rise to a stronger reduction of the entry field than surface roughness. This surprising result has been partially attributed to the fact that even for $H<H_p$ there will be vortices pinned into the inclusion (if its size is large enough) and will thus modify the current profile around the defect. The most salient prediction of that work is a universal supervelocity criterion for the vortex-entry into a type-II superconductors, which is shown to be more precise than the conventional local depairing current density criterion. It is shown indeed that the local current density at the corners of the defects is significantly less (by a factor 2 or 3) than the depairing current, due to the local suppression of the order parameter.

It is worth mentioning that a somewhat similar problem is that concerning the nucleation of vortices in superfluid $^4$He. The effect of a surface defect on the stability conditions of a vortex-free superflow of liquid $^4$He was treated by Soininen and Kopnin \cite{Soininen94} within the GL theory.

Further GL calculations and experimental investigations have addressed the effect of surface defects on the vortex expulsion and penetration in mesoscopic superconducting samples where confinement effects (and thus the shape of the sample) \cite{Baelus04,Vu18,Vu19,Embon,Alstrom10} as well as the exact location of the defects~\cite{Kuroda10,Vodolazov-2015} play an important role. In particular, it was shown that defects at the corner of a square shaped mesoscopic sample have little influence on the onset of vortex entrance, whereas defects located at the mid point of one side of the square lead to a clear reduction of the entry field \cite{Kuroda10}. In a recent work, Pack et al.~\cite{Pack2019} have extended the previous analysis concerning surface roughness to include local variations of the critical temperature for two and three dimensions. These local variations of the superconducting properties are notoriously relevant in Nb$_3$Sn cavities showing Sn segregations at grain boundaries~\cite{Carlson}. These authors show that even very small (on the order of the coherence length) surface roughness and material inhomogeneity can change the nucleation mechanism. The authors of Ref.~\cite{Babaev} formulated a method to compute the minimum energy path in a gauge theory, and applied it to solve the full nonlinear problem of vortex nucleation in the Ginzburg-Landau mode. They find that surface roughness at the scale of coherence length does not significantly affect the barrier for vortex nucleation.

In Ref.\cite{Vodolazov03b} it was shown that for an aluminium ring with a defect, the suppression of the critical field was 50\% in comparison with a nondefective ring. In Ref.\cite{Vodolazov05} it is indicated that for a defect of length $L$ much larger than the coherence length $\xi$ and of width $w\sim \xi$, the critical current density should decrease as $\sqrt{\xi/L}$. Note that in all cases described above, either using London or GL formalisms, pinning sites in the superconductor are not taken into account. For a recent investigation via time-dependent GL equations of the interplay between bulk and surface pinning mechanisms in a mesoscopic superconducting strip, see Ref. \cite {Kimmel18}. 

\subsection{Asymmetric distribution of border defects}

\begin{figure}[!ht]
\centering
\includegraphics*[width=1\linewidth]{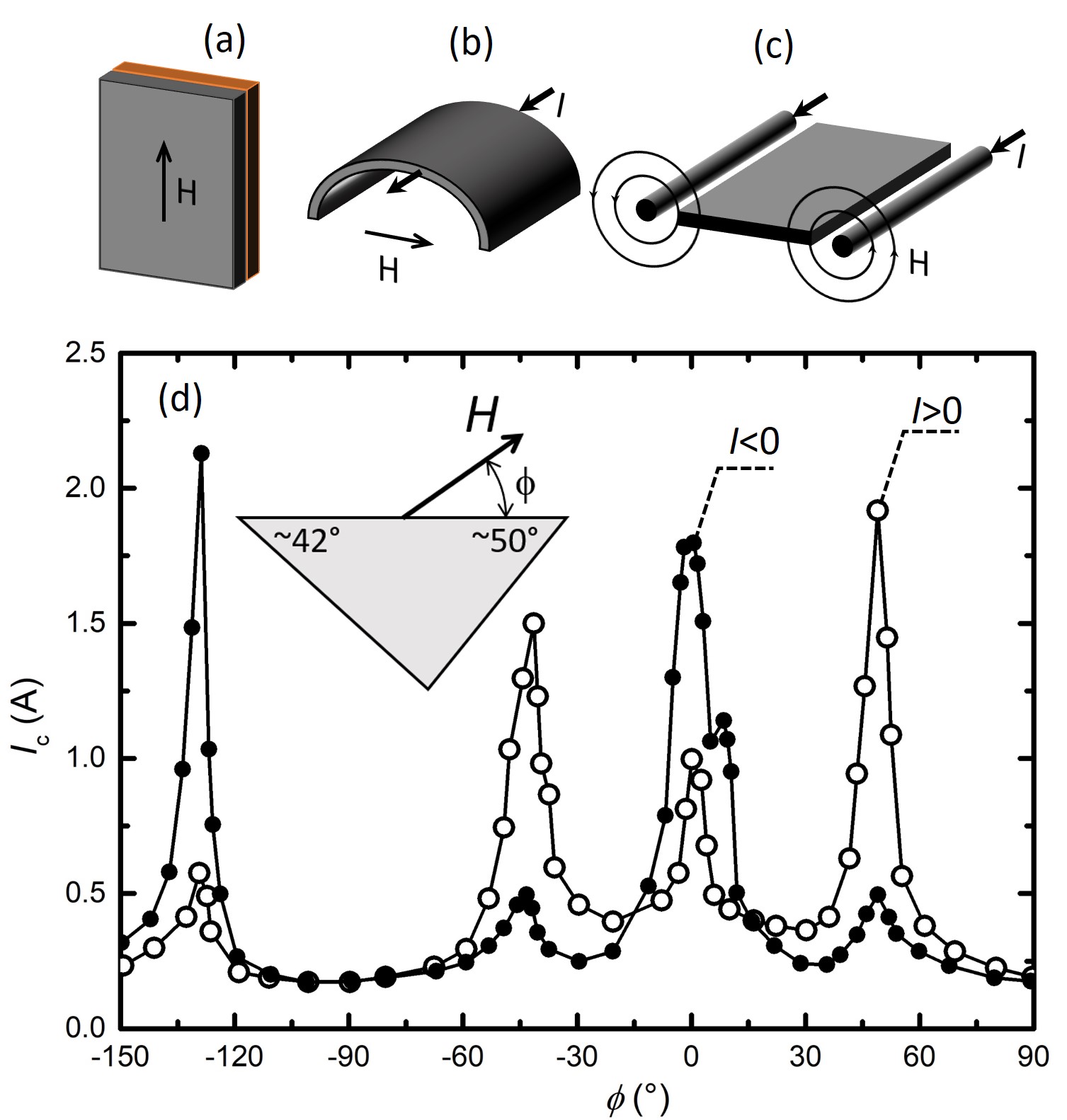}
\caption{Different configuration to induce rectification effects. (a) A bimetallic strip of Pb-Pb$_{0.9}$Tl$_{0.1}$ exhibits two different surfaces for magnetic flux penetration. (b) A superconductor with a curved surface under an applied magnetic field parallel to the cord of the arc. (c) A small perpendicular magnetic field is produced in a central superconducting strip by two wires laid parallel to the ribbon at its edges through which an applied control current is equally divided. (d) The critical current at 4.2 K of a well-annealed and polished prism of Pb$_{0.95}$Tl$_{0.05}$ as a function of the angle $\phi$ at a magnetic field of 700 Oe. The results show that the critical (surface) currents of opposite polarities go through maxima of significantly different magnitudes when the magnetic field vector is aligned parallel to one of the three surfaces. The current direction is perpendicular to the page. Adapted from \cite{Swartz67}.}\label{Swartz}
\end{figure}

Interestingly, the geometrical barrier delays only the penetration of flux but not its exit, thus imposing a natural asymmetry in the process of injecting and removing magnetic flux. The Bean-Livingston surface barrier delays not only the vortex entry but also the vortex exit, although it tends to exhibit a smaller barrier for the latter. This asymmetry comes from the fact that the magnetic flux $B$ inside the sample is nearly zero for vortex entry, whereas it has a constant finite value when starting to decrease the applied field $H$ after reaching $H>H_p$. The source of the asymmetry between entering and exiting is detailed in Refs.~\cite{Ternovskii,Clem1974,Burlachkov-1993, Burlachkov-1996}. This difference between vortex nucleation and vortex exit leads to hysteretic magnetization curves with vanishing magnetization at the descending branch of the magnetization loop due to the disappearance of surface barrier for flux exit at $H=B$ \cite{Konczykowski91,Geim97,Deo99}. Concerning electric transport properties, the impact of artificial surface modification on the onset of dissipation has been addressed as early as in 1967 by Swartz and Hart\cite{Swartz67}. These authors demonstrated experimentally that the critical surface transport current changes in magnitude when the direction of the transport current is reversed (rectification effect) for different cases of breaking the symmetry between entry and exit surface barriers. Examples of asymmetric surface rectifiers are shown in Fig.\ref{Swartz}. In Fig.\ref{Swartz}(a) a sandwich of Pb-Pb$_{0.9}$Tl$_{0.1}$ exhibits two different surfaces for magnetic flux penetration. In panels (b) and (c), the component of the local magnetic field normal to the surface has a different polarity on opposite sides of the superconducting stripe, thus breaking the symmetry and inducing rectification. In panel (d), this effect is achieved using a prism-shaped superconductor. Recently, similar ideas were implemented to manufacture micron-scale 3D superconducting diodes \cite{Moll} and investigate the diode effect in wedge-shaped superconductors \cite{Aguirre}. Another approach to induce an asymmetric surface barrier involves depositing a magnetic layer on top of a superconducting strip \cite{Vodolazov-2009}. 

\begin{figure}[!ht]
\centering
\includegraphics*[width=1\linewidth]{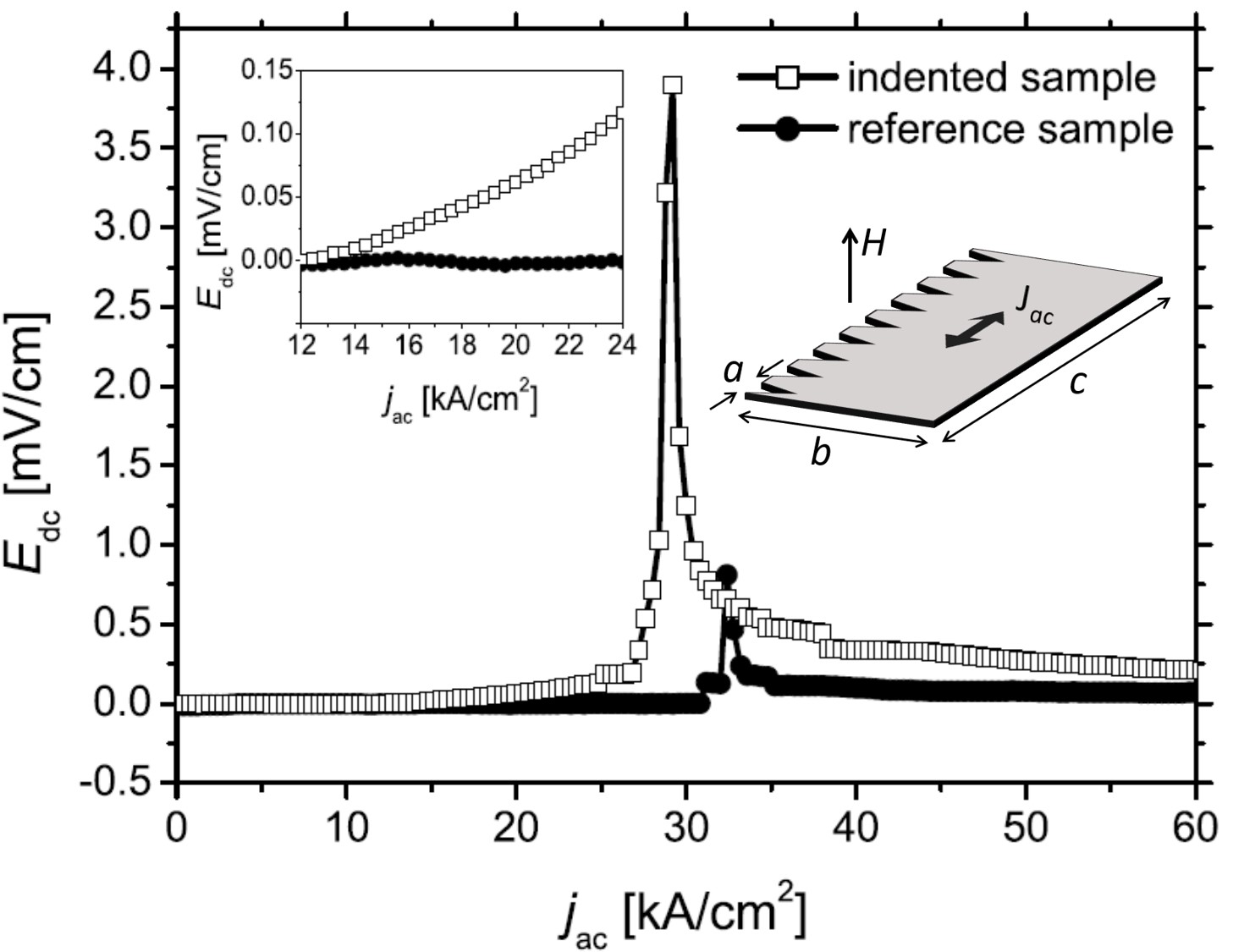}
\caption{Comparison of the rectified electric field $E_{dc}$ as a function of the ac excitation $j_{ac}$ for a reference symmetric strip and an indented (asymmetric) samples. The left inset shows a zoom for low current density values at the onset of the resistive regime. Measurements were done in a magnetic field perpendicular to the film surface $H = 0.55$ mT and for a driving frequency $f = 33.711$ kHz. The cryostat temperature during measurements was kept at $T = 1.12$ K. Rightmost inset: Layout of the sample investigated in Ref.\cite{Cerbu13} with $a = 2.5 \mu$m, $b = 50 \mu$m, $c = 200 \mu$m and thickness $t = 50$ nm. Figure adapted from \cite{Cerbu13}.}\label{Cerbu-1}
\end{figure}

A superconducting rectifier based on the asymmetric surface barrier effect was theoretically proposed and investigated by Vodolazov and Peeters \cite{Vodolazov05}. Here, by roughing only one of the edges parallel to the current, the conditions for vortex entry become different on both edges of the sample. Later on, the impact of a microfabricated artificial roughness at the border of a superconducting thin film was experimentally investigated by Cerbu \etal \cite{Cerbu13} by transport measurements and complemented with time-dependent Ginzburg-Landau simulations. The geometry of the investigated Al sample shown in the rightmost inset of Fig.\ref{Cerbu-1} consists of a 50 $\mu$m wide strip in which one of the borders has been patterned in saw-tooth shape with individual indentation size $a=2.5~\mu$m whereas the opposite side of the strip contains a natural roughness of about 0.25 $\mu$m. 

The authors of Ref.\cite{Cerbu13} applied an ac current of zero mean value, in such a way that if both borders are identical, a zero DC voltage (or electric field $E_{dc}$) should be detected. In Fig.\ref{Cerbu-1} a comparison of the resulting  $E_{dc}$ as a function of the ac excitation $j_{ac}$ is shown for a nominally symmetric sample and a sample with asymmetric borders. The sample with symmetric borders shows a weak signal, whereas the indented sample exhibits a strong rectification effect. The origin of the rectification signal comes from the fact that what defines the critical current is vortex-entering into the sample rather than vortex exiting it.

\begin{figure}[!ht]
\centering
\includegraphics*[width=1\linewidth]{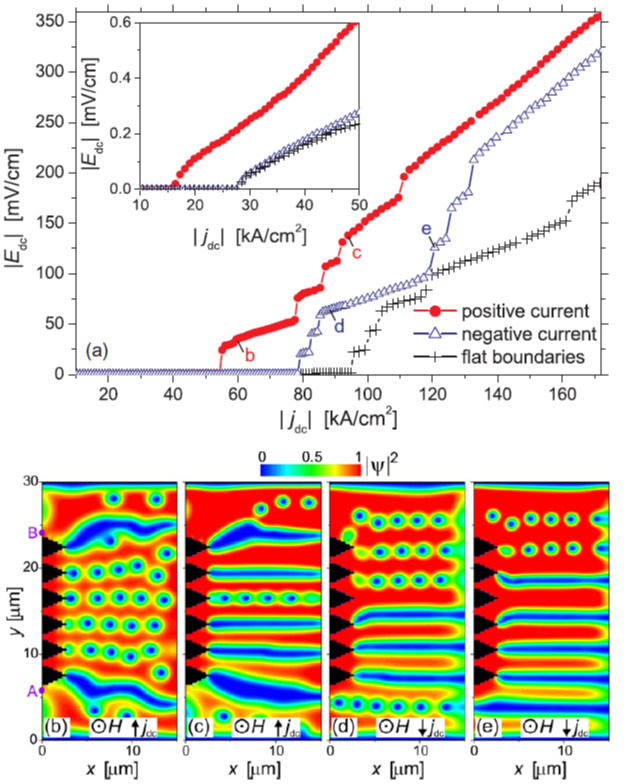}
\caption{(a) Absolute value of the calculated average electric field $E_{dc}$ between points $A$ and $B$ (see panel (b)) as a function of the increasing density $|j_{dc}|$ of the applied dc current for positive (filled circles) and negative (open triangles) current directions. For comparison, the results for a symmetric bridge with flat boundaries are shown with crosses. Snapshots of the order parameter distributions, which correspond to points labeled as `b' ($j_{dc}$ = 59 kA cm $^{-2}$), ‘c’ ($j_{dc}$ = 94 kA cm$^{-2}$), ‘d’ ($j_{dc}$ = -87 kA cm$^{-2}$) and ‘e’ ( $j_{dc}$ = -122 kA cm$^{-2}$), are plotted in panels (b)-(e), respectively. The inset of panel (a) shows the behavior of $E_{dc}$( $j_{dc}$) at relatively low current densities.
  Figure reproduced from \cite{Cerbu13}.}\label{Cerbu-3}
\end{figure}

Time-dependent Ginzburg-Landau simulations for a sample of similar geometry and size as in the experiment are shown in Fig.\ref{Cerbu-3}. Panel (a) shows the calculated time-averaged electric field $E_{dc}$ between points $A$ and $B$ indicated in panel (b), as a function of the current density $j_{dc}$ of positive (filled circles) or negative (open triangles) polarity. For positive polarity, the onset of dissipation takes place at a smaller $j_{dc}$ than for negative polarity. This result is expected since the indentations greatly facilitate the nucleation of vortices and their entry into the superconductor. Even at current densities quite above the onset of the resistive regime in the case of flat boundaries, vortex entry on the micropatterned boundary takes place exclusively through the vertices of indentations. This is illustrated in Fig.\ref{Cerbu-3}(b) with a plot of the distribution of the streaming parameter $S=\left[ (t_2-t_1)^{-1} \int_{t_1}^{t_2}(\partial |\psi|^2/\partial t)^2 dt \right]^{1/2}$ introduced to visualize the vortex trajectories. In panel (a),  the results calculated for a symmetric strip with flat boundaries at both sides (crosses) are also shown. As follows from the inset of panel (a), in this case the magnitude of the critical current, which corresponds to the onset of the resistive regime, is practically the same as that for negative currents in the strip with indentations. This clearly demonstrates that the onset of the resistive regime is mainly determined by the possibility for vortices to enter the superconductor and hence by the properties of the ‘inlet’ boundary of the strip, rather than by the properties of the ‘outlet’ boundary. Similar effect has been reported in a Cu/MoN strip with a single cut near one of the strip edges \cite{Ustavschikov-2022}. The observed diode effect, in principle able to operate up to GHz frequency, was associated with the condensation of current lines near the cut, which leads to different conditions of entrance of vortices from the strip edge with the cut and from the opposite edge without a cut.

\begin{figure}[!ht]
\centering
\includegraphics*[width=1\linewidth]{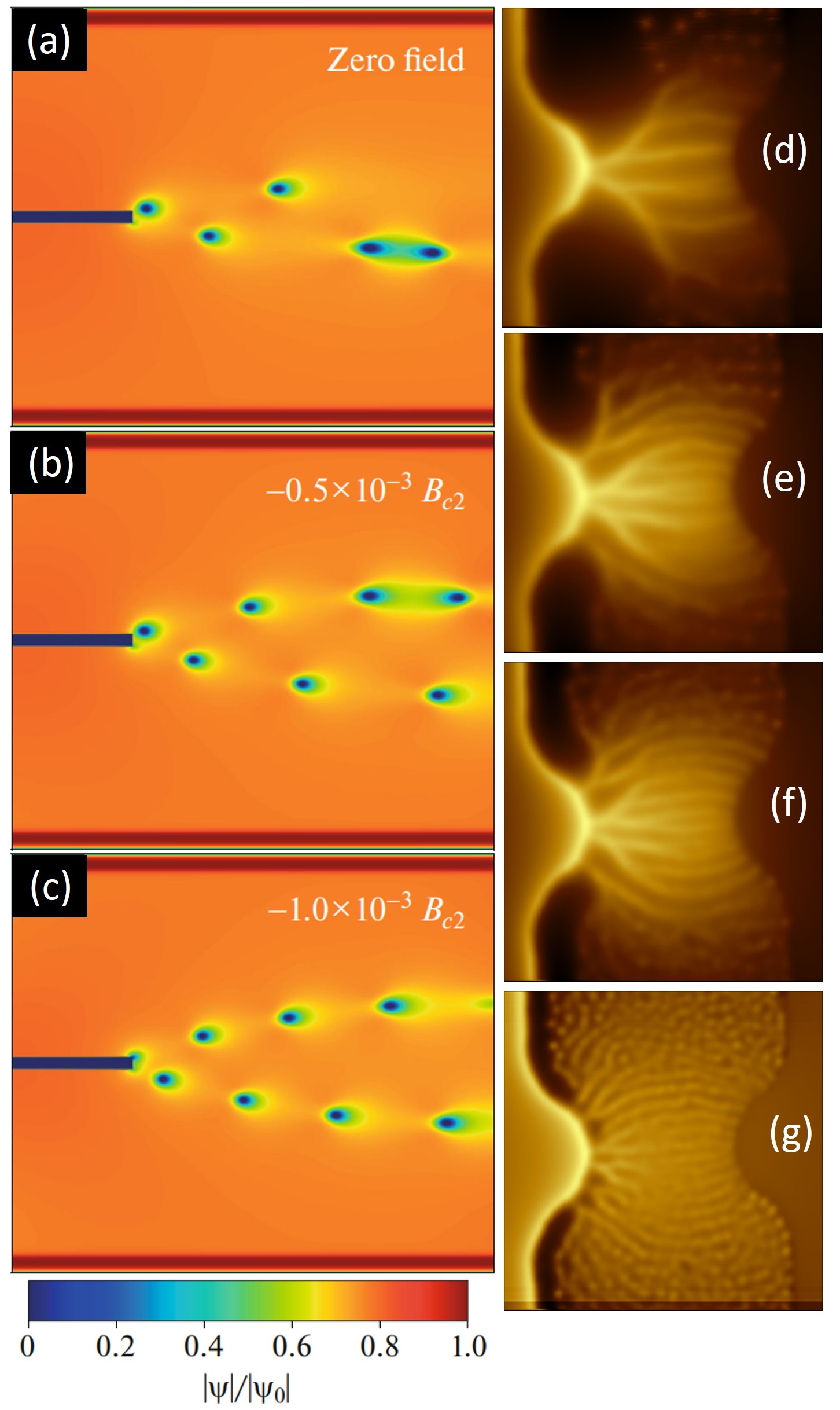}
\caption{(a-c) Simulated spatial distributions of the superconducting order parameter $|\psi|$ at a random instant in a superconducting strip with a transverse cut. Dark blue regions opposite to the cut show the normal cores of vortices. Since the order parameter relaxes within a finite  time, for vortices at high velocities, the order parameter has no time to recover during the vortex time of flight. In such a case, regions with a suppressed order parameter (light-yellow and blue regions) appear between the cores of vortices. Adapted from Ref.~\cite{Ustavschikov-2023}. (d-g) Experimentally imaged vortex trajectories using SQUID-on-tip technique at 4.2 K for a current and magnetic field of 21 mA and 2.7 mT (d), 18 mA and 4.2 mT (e), 16 mA and 5.4 mT (f), and 11.8 mA and 9 mT (g) in a Pb film with a 5.7-µm-wide central constriction. Adapted from ref.~\cite{Embon}.}\label{Menorah}
\end{figure}

Along the same vein, Budinska et al.\cite{Budinska-2022} investigated the development of fast vortex motion, so-called flux flow instabilities, in 15-nm-thick 182-$\mu$m-wide MoSi strips with rough and smooth edges produced by laser etching and milling by a focused ion beam. For the strip with smooth edges, a critical current three times larger and a factor of 20 higher maximal vortex velocity ($\sim$ 20 km/s) compared to the rough border, were found. Interestingly, in a later work, Bezuglyj et al.\cite{Bezuglyj-2022} provided a comprehensive description of the trajectory followed by vortices penetrating from an edge indentation under the influence of an applied current. These authors propose an elegant approach to detect the velocity component of these vortices along the direction of the applied current and, based on this, determine the formation of vortex jets. The vortex jets are narrow near the defect and expand due to the repulsion of vortices as they move to the opposite edge of the strip. In a follow-up work, Bevz et al.\cite{Bevz-2023,Bevz-2023b} working on MoSi strips and simultaneously Ustavschikov et al.\cite{Ustavschikov-2023} on MoN strips, proposed a method to unambiguously determine experimentally the exact number of vortices traversing the strip. By carving a very narrow slit perpendicular to the current flow direction, step-like structures develop in the voltage-current characteristics, which were attributed to a step-wise increase in the number of vortices simultaneously traversing the bridge. These results were compared to time-dependent Ginzburg-Landau numerical modeling and Aslamazov-Larkin model \cite{Aslamazov-Larkin-2}. Illustrative examples of the simulated and measured dynamic vortex patterns emerging from a narrow slit and a lateral constriction, are shown in Fig.\ref{Menorah}.

Interestingly, when the MoN strips with a side cut are exposed to microwave irradiation, Shapiro-like steps and regions with a negative differential resistance were observed in the voltage–current characteristics \cite{Ustavschikov-2022b}. The former are particularly pronounced at a low power of microwave radiation, and the latter at high-power irradiation. Based on time-dependent Ginzburg–Landau simulations and heat conduction equations for the electron temperature, the authors suggest that the negative differential resistance is due to the disordered (chaotic) motion of vortices across the strip near the cut. Interestingly, in Refs.~\cite{Reichhardt1997,Gutierrez2009}, a similar turbulent to laminar-like vortex flow transition was observed in an Al film with a periodic array of pinning centers, which also manifested as a negative differential resistivity. A somewhat related physics is discussed in Ref.\cite{Vodolazov-2015} concerning the effect of a single circular hole on the critical current of a narrow superconducting strip with width $W$ much smaller than Pearl penetration depth. When the hole sits close to the edge of the strip, the current-generated vortices enter the hole via the edge of the film, and then they move across the superconductor, resembling the effect of an edge defect. When the hole is located away from the edges, a vortex-antivortex pair is nucleated inside the hole and a motion of the pair destroys the superconducting state. Depending on which one of the two mechanisms prevails, it results in a very distinct nonmonotonic dependence of critical current on the position of the hole. The case of a a mesoscopic superconducting strip with an engineered defect at the center in which the defect is another superconductor with a different critical temperature, has been investigation via TDGL simulation in Ref.~\cite{Barba-Ortega}. The authors considered several geometrical shapes of the defect and found that for a non-centered square defect, the vortex-antivortex pair nucleates nonsymmetrically in the sample due to the asymmetry of the supercurrent in the sample, and that a triangular defect breaks the spatial symmetry for vortex-antivortex motion. Remarkably, we are nowadays witnessing a renewed interest in the non-reciprocal vortex displacement in the context of vortex diode effect \cite{moodera,clecio,Bergreen-2024,Ingla-Aynes2025}.

\subsection{Strategies for increasing the penetration field}

There is a genuine technological interest in understanding the mechanism by which vortices penetrate through surface defects into superconducting materials. This is particularly the case in superconducting linear accelerators consisting of 16000 units of meter-long superconducting cavities needed to drive the particle beams \cite{ILC} at resonances of 0.1-2 GHz and currently made of pure Nb or on oxygen-free copper (mm thickness) coated with an Nb layer (µm thickness). These systems are immersed in a pumped $^4$He liquid bath below the superfluid transition. The advantage of using superconducting resonators compared with normal metals is that dissipation (quantified by the surface resistance) is orders of magnitude lower in superconducting materials, and, therefore, quality factors (i.e. the ratio of stored electromagnetic energy to mean dissipated power) as high as 10$^{11}$ can be achieved. A high-quality factor implies the possibility of applying a high accelerating electric field, which in turn corresponds to high RF magnetic fields parallel to the Nb surface. These high Q values are only possible if the cavity remains in the Meissner state. This is particularly challenging since cooling down the cavity in the presence of the Earth's magnetic field leads to a rapid deterioration of the cavity performance due to the trapped vortices in the material defects. Note that these trapped vortices are not necessarily parallel to the surface and reach high velocities when shaken by the RF field. Reducing the local magnetic field by effective screening with Mumetal, for instance, does not fully eliminate the presence of vortices. 


A diversity of strategies with the aim to improve the quality factor (or reduce the surface resistance) of superconducting cavities have been investigated. For instance, it has been shown that spatial temperature gradients can act as thermal brooms able to decrease the trapped flux and, thus, the residual resistance. This is achieved via thermal cycling protocols as discussed in Ref.~\cite{vogt1,vogt2}. Experiments carried out to explore the expulsion of magnetic flux by thermal broom \cite{Romanenko2014,Posen2016,Romanenko2014} show that the larger the temperature gradient, the more efficient is the removal of the penetrated vortices. This has been confirmed with numerical simulations using time-dependent Ginzburg-Landau (TDGL) theory by Li et al. \cite{He2023-1} and He et al. \cite{He2023-2}. An alternative way to reduce the low-field surface resistance (and thus improve the quality factor) is to lower the mean free path of the material by nitrogen-doping \cite{Gonnella}. Further optimizations of flux pinning to reduce flux sensitivity at low RF fields in niobium superconducting radio-frequency cavities are discussed in Ref.\cite{Dhakal}. As a matter of fact, the penetrated field depends not only on the disorder in the vicinity of the surface, but also on the geometry of the vacuum-superconductor interface. For a planar half-space geometry, the effects of homogeneous disorder and engineered multilayer superconductor-insulator structures have been investigated  \cite{Lin-2012,Kubo2014,Posen2015,Gurevich2015,Liarte2016,Liarte2017,Kubo2017}. It was found that the homogeneous disorder increases the penetration depth but reduces the critical current, leading to a modest enhancement of the superheating field at low temperatures. The inhomogeneous disorder introduced by injecting nitrogen into the surface layer of the niobium SRF cavity has been found to increase the superheating field above the maximum value of the superconductor's clean limit or uniformly distributed disorder \cite{Ngampruetikorn2019}.

Numerical simulations have been shown to be instrumental in understanding the complex physics of vortex penetration and RF shaking dynamics in superconducting cavities, while also providing a pathway to identify mitigation strategies \cite{Carlson,Wang-2022,Kimmel18,Kubo-Gurevich,Xue-review-2024}. A remarkable example is the proposition of cladding a Nb cavity with a thin insulating layer and a superconducting layer thinner than the superconducting penetration depth $\lambda$ and having a higher critical temperature than Nb \cite{Gurevich-2006}. For a bilayer material consisting of a dirty superconductor on top of a clean superconductor, Ref. \cite{Kubo2017} reported that the superheating field can exceed that of either material. The superheating field may also be increased by introducing insulating layers to retard flux line penetration \cite{Kubo2014,Posen2015,Liarte2016}. On the same vein, a multilayer comprised of alternating thin superconducting and insulating layers on a thick substrate can very efficiently screen the applied magnetic field exceeding the superheating fields of both the superconducting layers and the substrate \cite{Gurevich2015,Wang-2022,Gurevich-2006,Pathirana2023}. Recently, Asaduzzaman et al.\cite{Asaduzzaman} confirmed via a µSR study that the introduction of a thin superconducting Nb$_3$Sn overlayer on Nb can effectively push the onset of vortex penetration up the superheating field. As expected, surface roughness can cause local field concentration and thus reduce the edge barrier, resulting in magnetic flux penetration at lower fields than the superheated field \cite{Porter2016}. As a consequence, a higher penetration field can be achieved by a smoother surface, as reported in Refs. \cite{Shemelin2008,Kubo2015,Pack2019,Xie2013}. Another approach deserving further investigation is the possibility of inducing enhancement of the Bean-Livingston barrier by coating the superconducting material with a soft ferromagnet, as discussed in Ref.\cite{Genenko}.

\section{Continuous media description: Magnetic flux pattern produced by border indentations}
\label{Sec3}

We now turn to analyzing the impact of defects on the penetration of
magnetic flux over macroscopic length scales, $\ell$, which are much
larger than the magnetic penetration depth, $\lambda$, or the vortex
mean separation distance, $\sim (\phi_0/B)^{1/2}$. Similar to the
microscopic picture, the extent to which a border defect influences
the magnetic flux distribution is intimately related to the 
perturbation of the electric current flow within the superconductor.

\subsection{Critical state approach}

\paragraph{Critical state model}
The simplest model that illustrates the ``brick wall'' effect of
Fig.~2 is the critical state model (CSM) ~\cite{Bean1962}, applied to
superconductors with strong pinning. The CSM assumes that upon an
external change of the magnetic flux, any region of the superconductor
being subjected to an electromotive force, however small, reaches a
critical state with a local current density, $\bf{j}$, which is
determined by a local static equilibrium between pinning forces and
the Lorentz-like force, $\bf{j}\times\bf{B}$ ~\cite{Campbell1972}. In
its simplest form, where $H_{c1}$ and surface barriers are neglected,
while $j_c$ is assumed to be independent of $\bf{B}$, the CSM is
written
\begin{eqnarray}
  \label{eq:CSM}
  \nabla\times\bf{H} = \bf{j},
\end{eqnarray}
where $|\bf{j}|=j_c$ in regions which have been subjected to a change
in magnetic flux, $|\bf{j}|$$=0$ in virgin regions, and $\bf{H}$ is in
equilibrium with the external field at the surface of the specimen. 

\begin{figure}[ht]
\begin{center}
  \includegraphics[width=0.25\textwidth]{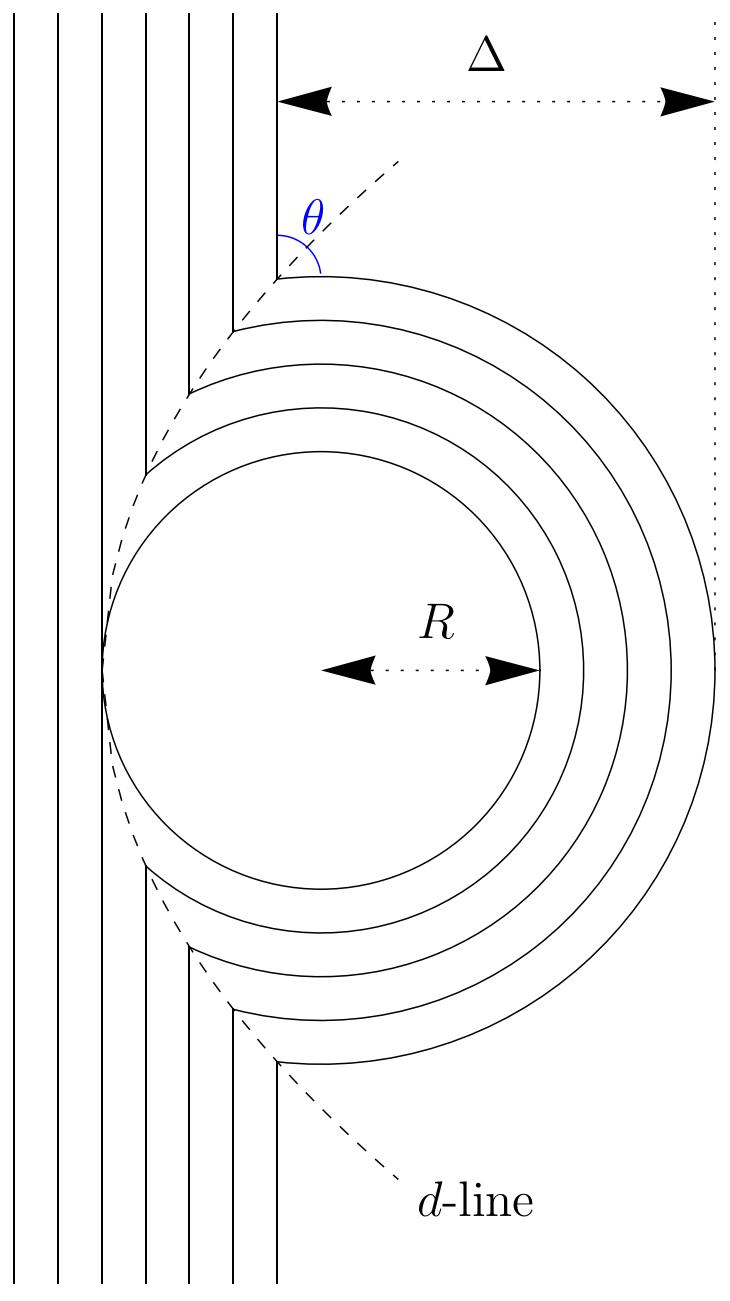}
  \caption{\label{f:hole-Campbell}Contours of the magnetic flux
      density around a cylindrical hole of radius $R$. The magnetic
      flux density is penetrating from the left side. The dashed line
      (the so called $d$-line) shows the locations where the direction
      of the induced current density is discontinuous. The angle
      $\theta$ (in blue) shows the total deflection angle of one of
      the current density lines (see text). As a result of the
      currents flowing around the hole, there is an excess penetration
      $\Delta = 2 R$ between the flux front ahead of the hole and that
      diffusing from the sample border.  Adapted from
      \cite{Campbell1972}.}
\end{center}
\end{figure}

The CSM magnetic flux distributions exhibit a certain ``rigidity'',
leading to patterns similar to the brick wall pattern of
Fig. 2. Consider first a long specimen containing a cylindrical hole
at some distance from a flat surface. Upon magnetization, the CSM
yields a flat flux front penetrating across the surface into the
interior of the specimen. In such long geometries, contours of
magnetic flux density coincide with the current density lines. A
uniform $j_c$ thus yields a steady decrease of the magnetic flux
density from the border to the front, with equally spaced current
lines. Upon reaching the hole, the magnetic flux invades the cavity
and the critical state forms around the perimeter, subsequently
generating the magnetic flux distribution of
Fig.~\ref{f:hole-Campbell}.  This distribution is characterized by a
discontinuity line, or $d$-line, marking an abrupt change in the
direction of $\bf{j}$. Here, the $d$-line takes the shape of a
parabola and extends to infinity. The equal spacing of current lines
thus leads to an undamped perturbation which extends to the entire semi-space.

Similar current-line patterns arise in flat film geometries. As a general rule, the solutions for the current density distribution found in long samples are also valid for thin samples in perpendicular fields and are good approximations in thin sample with a partial penetration~\cite{Schuster1994}. If the
film is assumed thin, the problem can be formulated in terms of the 2D
sheet current density, $\bf{J}$, which corresponds to the current
density $\bf{j}$ integrated over the film thickness,

\begin{eqnarray}
  \label{eq:sheet-density}
  {\bf J} = \int_{-d/2}^{d/2} {\bf j}\: dz,
\end{eqnarray}

\noindent where $d$ is the film thickness, and which can be determined from the
streamline function, $g$, as
${\bf J} = {\bf \nabla}\times (g \hat{z})$. The CSM solution of this
problem was established by Prigozhin~\cite{Prigozhin1998} through a
variational formulation. It was shown that $g(\bf{r})$ at position
$\bf{r}$ is proportional to the Euclidean distance from the boundary
of the specimen, $\Gamma$ to $\bf{r}$
\begin{eqnarray}
  g(\bf{r}) = J_c \,\mathrm{dist} (\bf{r},\Gamma). 
\end{eqnarray}
Here, $g$ thus plays a similar role as $\bf{H} = \bf{B}/\mu_0$ in
long specimens. More specifically, a circular hole in a thin film
produces the same current line pattern as in
Fig.~\ref{f:hole-Campbell}.

\begin{figure}[ht]
\begin{center}
  \includegraphics[width=0.5\textwidth]{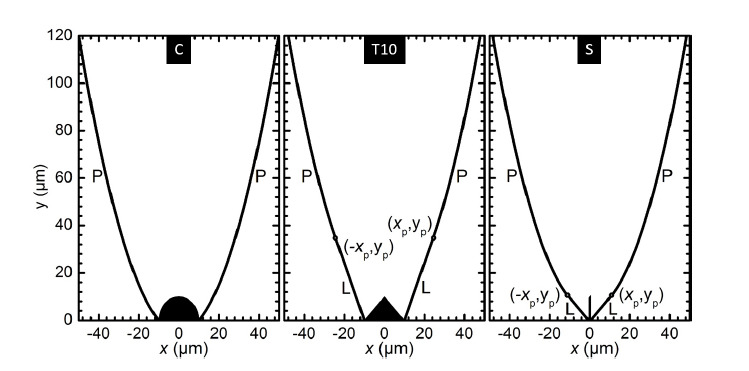}
  \caption{\label{f:Bean-shapes}Discontinuity lines predicted in the
    CSM model for a semicircular, a triangular, and a thin
    rectangular indentations, respectively. The $d$-lines are described by parabola branches P, but for the triangular and slit indentations, the branches in the vicinity of the defect ($y < y_P$) correspond to straight lines L. Reproduced from Ref.\cite{Brisbois2016}}
\end{center}
\end{figure}

The overall shape of the $d$-lines extending away from the defect is a
signature of the size and shape of the defect, as illustrated
in Fig.~\ref{f:Bean-shapes}, showing $d$-lines generated in samples with
straight edges containing respectively a semicircular (labelled C), a triangular (labelled T10),
and a rectangular (labelled S) indentation. As discussed above, the discontinuity
lines of Fig.~\ref{f:Bean-shapes} arise in either long samples (where
$|{\bf j}|=|{\bf \nabla}\times {\bf H}|={\bf j_c}$ in the flux penetrated region)
or in thin films (where $|{\bf J}| = |{\bf \nabla} \times g\hat{z}|={\bf j_c} d$
in fully penetrated regions). A semicircular indentation of radius $R$
is seen to produce two arcs of a parabola, with
\begin{eqnarray}
  \label{eq:semicircular}
  y(x) = \frac{x^2}{2 R} - \frac{R}{2},
\end{eqnarray}
which has a radius of curvature
$R_0 = \left.d^2y/dx^2\right|_{x=0} = R$, thus directly
related to the actual size of the defect. The relationship between the
$d$-lines and the defect size is less direct for the triangular and
rectangular indentations. As discussed in Ref.~\cite{Brisbois2016}, the
$d$-lines typically start near the defect in a way that strongly
depends on the defect geometry, whereas they extend into the film
asymptotically following the branches of a parabola with $y \sim x^2/2 R$.
The separation of the branches is directly related to the width of the
defect along the edge. The second derivative of the branches, on the
other hand, is given as $d^2y/dx^2 \simeq R$ where $R$ is the
extension of the defect perpendicular to the straight
edge.

For the case of thin films, an additional element of information is
contained in the intensity of the discontinuity lines, as observed for
instance in magneto-optical images. Near the $d$-line, the magnetic flux
density exhibits a logarithmic discontinuity with
\begin{eqnarray}
  \label{eq:contrast}
  B_z = \pm K_l\,\frac{\mu_0}{2\pi}\,j_cd\,\ln(\delta x),
\end{eqnarray}
where $\delta x$ is the distance from the $d$-line,
$K_l = 2\cos(\theta/2)$ is a contrast factor, and $\theta$ is the
total deflection angle of the current lines at the bending
point~\cite{Schuster1994} (This angle is shown in blue for one of the current lines in Fig.~\ref{f:hole-Campbell}). The contrast
factor simply expresses the fact that a strong bending of the current
flow at the $d$-line locally enhances the magnetic flux density and
hence makes the $d$-line appear sharper on the magneto-optical
image. For curved lines such as a parabola arising near a circular
defect, the contrast factor varies continously along the $d$-line, so
that the parabola appears sharp near its apex, while it progressively
blurs out at larger distances.

Generalizing now to defects of arbitrary shapes, the current flow is
characterized by discontinuity lines taking their origin in convex
corners, whereas the current lines flow continuously around concave
corners~\cite{Schuster1996}. At far distances from the tip of a
concave corner, the discontinuity lines are defined as the location of
equal distances between the tip and the border and hence form branches
of a parabola. Last, the $d$-lines shape are dictated by the geometry of
the sample: whereas large samples with straight edges exhibit
typically parabolic $d$-lines at a sizable distance from the defect,
circular samples show a $d$-line pattern which closes on itself and
forms an approximately elliptical shape~\cite{Jooss1998a}.

\begin{figure}
  \includegraphics[width=\linewidth]{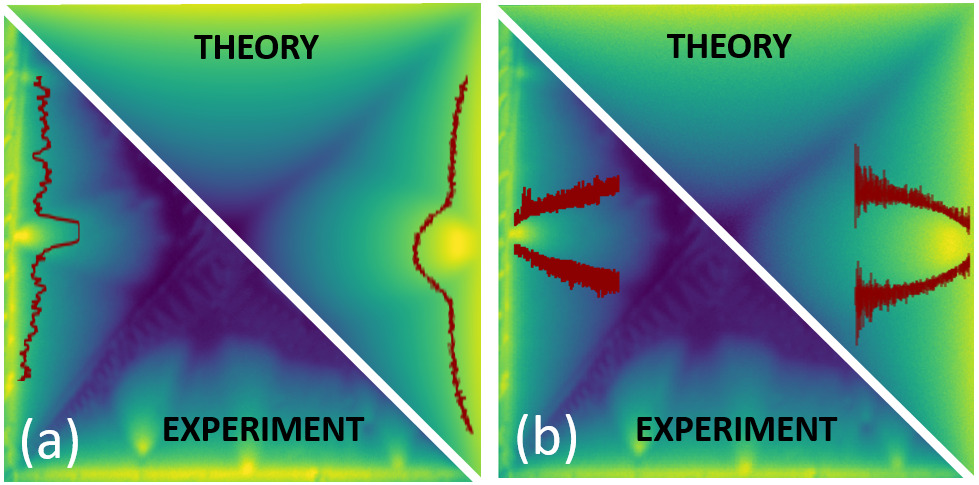}
  \caption[$d$-lines: experiemental vs numerical
  computation]{Comparaison of experimental MOI data with numerical
    simulations with added Gaussian noise for YBCO thin
    films. \textbf{(a)} Intensity profiles of a pixel line parallel to
    the sample edge. \textbf{(b)} Regions containing the $d$-line as
    given by the image analysis. Reproduced from Ref.\cite{S-Blanco-thesis}.}
  \label{fig:dline profile}
\end{figure}

In a series of experiments, Blanco Alvarez \emph{et al.} confronted the
predictions of the critical state model to magneto-optical data of the magnetic
flux penetration in YBCO square films with artificial indentations of
different shapes~\cite{S-Blanco-thesis}. The samples were 100~nm thick
square films of 800~$\mu$m side, grown epitaxially on a LaAlO$_3$
substrates by chemical solution deposition. The indentations were made
by laser direct writing on a photoresist and dry etching with ion
milling, in various polygonal shapes. Discontinuity lines were mapped
out by a careful image analysis based on locating sharp variations of
the magnetic field, while correcting for inhomogeneities in the light
source and reducing the image noise. The same procedure was applied on
numerical simulations of the MOI images performed at an elevation of
5~$\mu$m above the YBCO films, accounting for the gap between the
magneto-optical indicator and the sample, to which a Gaussian noise
was added to emulate the observed experimental noise. The results presented in Fig.~\ref{fig:dline profile} for a $8.5~\mu$m$\times 10~\mu$m
rectangular indentation, show a reasonable agreement between
experiment and theory. For such an indentation, a parabolic shape is
expected for distances greater than the indentation depth and the
corresponding contrast factor is expected to decrease along the
$d$-line, as explained above. It is worth noting that the loss of
contrast coincides with the increase in the size of the error bar,
thus, the latter can be used to estimate the contrast factor. A total
of six different shapes were investigated and compared to the
predictions of the critical state model, as shown in Fig.~\ref{fig:dline
  geometry}. Overall, the agreement between theory and experiments is
very good.

\begin{figure}
  \includegraphics[width=\linewidth]{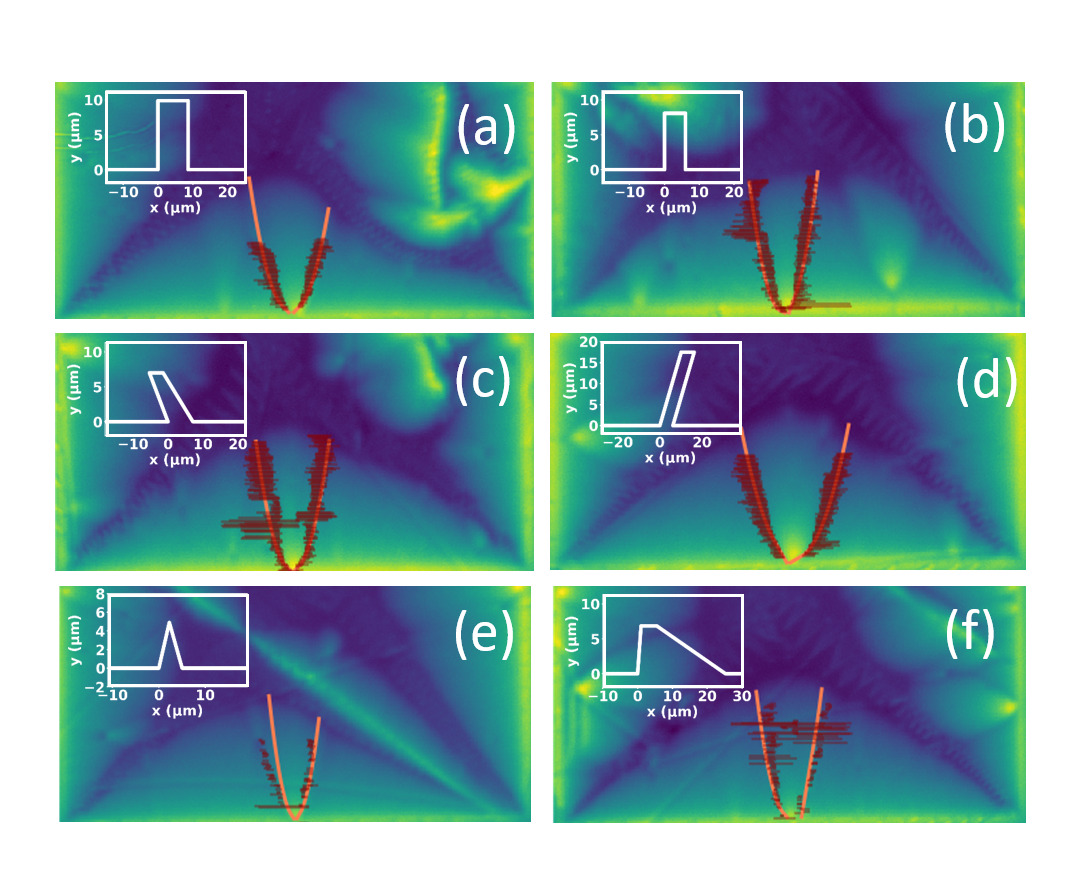}
  \caption[Influence of indentation shape on $d$-lines]{Influence of
    indentation shape on $d$-lines for various indentations. The shape
    of each indentation is represented in white. The picture
    represents magneto-optical images of the magnetic landscape under
    $\mu_0 H = 10$ mT at $T/T_c=$ 0.69. The blue regions correspond to
    low magnetic field regions while the green-yellow regions are
    those with the highest magnetic field values. The salmon colored curves
    show the predictions of the critical state model for the different
    indentations, while the maroon marks show the numerically detected
    $d$-lines. Reproduced from Ref.\cite{S-Blanco-thesis}.}
  \label{fig:dline geometry}
\end{figure}

\begin{figure}[ht]
\begin{center}
  \includegraphics[width=0.5\textwidth]{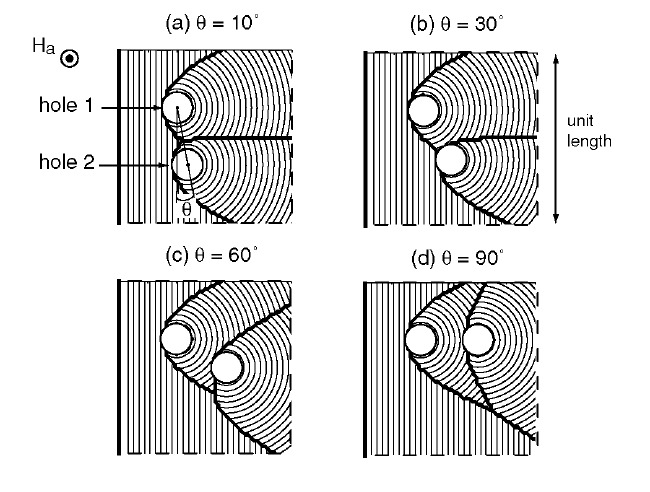}
  \caption{\label{f:array-defects}Network of $d$-lines generated in a
    long specimen with two cylindrical holes, from Ref.~\cite{Lousberg2008}.}
\end{center}
\end{figure}

The current perturbation due to several defects exhibits yet more
complex patterns. Two defects placed side by side along the
superconductor border share a common $d$-line extending perpendicular
to the border, whereas an arbitrary placement of the defects
gives rise to specific networks of $d$-lines, as illustrated in
Fig.~\ref{f:array-defects}. It has been shown that macroscopic drilled holes in high-Tc
superconductors, introduced with the aim to improve oxygen
diffusion and heat exchange, need to be placed strategically
along the $d$-lines generated by neighboring holes if trapped
magnetic flux is to be maximized~\cite{Lousberg2008}.

\paragraph{Border defects due to inhomogeneities}
So far, only non-conducting defects such as holes and indentations
were considered, and it was shown that the generated discontinuity
lines carry useful information about the defect geometry. It is
interesting to extend the analysis to other types of defects, such as
a region with a reduced critical current density, and determine
whether similar discontiuity lines are generated and what information
can be drawn from their experimental observation. In an early work,
Schuster \emph{et al.}  investigated the influence of an inhomogeneous
distribution of critical current density on the penetration of the
magnetic flux in rectangular films~\cite{Schuster1995}. The variation
in the critical current density was obtained by thinning the
superconducting film in predefined regions.  The resulting critical
current density varied abruptly and gave rise to a network of
  $d$-lines that can be easily obtained by requiring a conservation of
  the normal component of the current density. In addition, it was
  predicted that the boundaries separating regions of different
  critical current densities are crossed by magnetic flux during
  penetration. As a result, an electric field is generated along these
  boundaries and can be strongly enhanced at the points where the flux
  transfer is the highest. Inhomogeneities in the current densities
have also been considered in the context of flux avalanches in the
presence of defects or variations of the superconduting properties,
either arising during the material
synthesis~\cite{Welling2004,Solovyov2013}, or introduced artificially
~\cite{Motta2014}, as discussed in the next Section. The inhomogeneity
of the critical current distribution was also shown to strongly affect
both the generation and the subsequent motion of flux
avalanches~\cite{Albrecht2007, Treiber2010, Treiber2011, Jing2016,
  Lu2016}. More specifically, it is thought that avalanches branching
is caused by local variations of the critical current
density~\cite{Aranson2005}.

\begin{figure}[ht]
\begin{center}
  \includegraphics[width=0.3\textwidth]{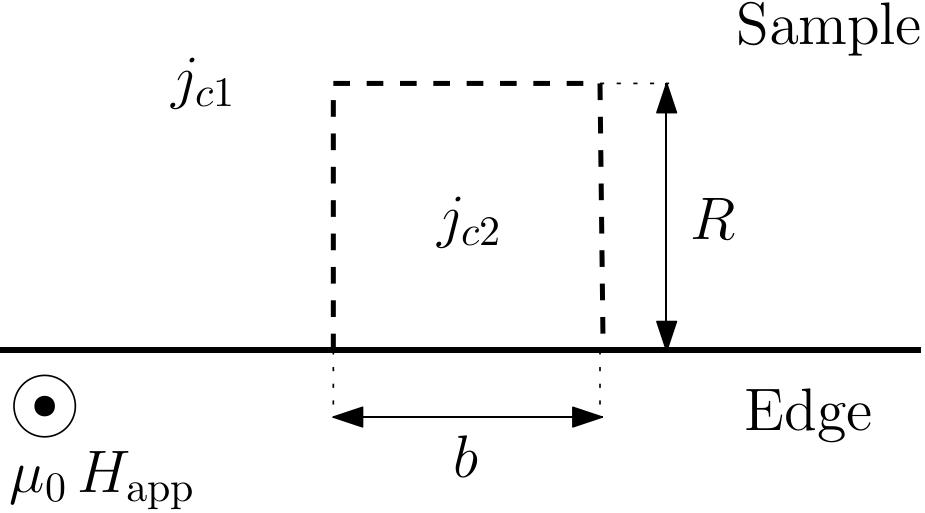}
\end{center}
\caption{Elementary model for a defect due to an inhomogeneity in
    the sheet current density. A region of size $b \times R$ along the edge of the sample has a critical current density $j_{c2}$ different from the rest of the sample, of critical current density $j_{c1}$. The sample is subjected to a perpendicular magnetic flux density $\mu_0\,H_{\mathrm{app}}$.}
  \label{f:defect-inhomogeneous}
\end{figure}

\begin{figure}[ht]
\begin{center}
\includegraphics[width=0.5\textwidth]{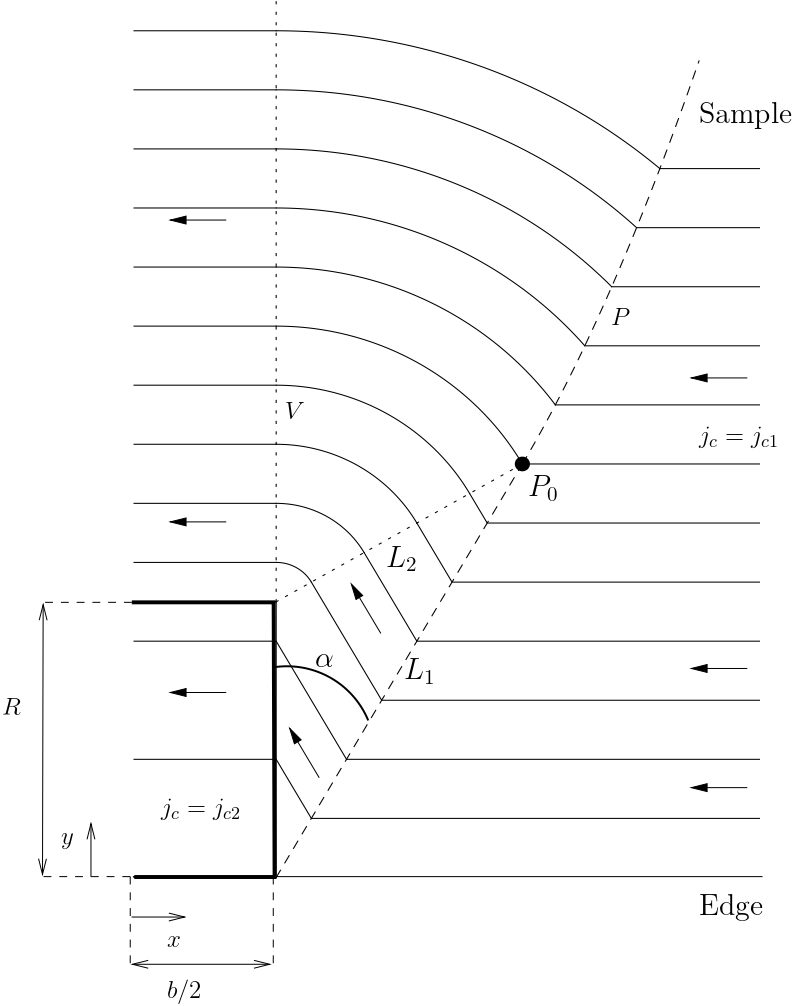}
\end{center}
\caption{Current distribution for the case of a higher current density
  outside the defect. The $d$-lines $L_1$ and $P$ are depicted with
  dashed lines. The dotted lines $V$ and $L_2$ are shown to reveal the
  current line structure. They delimit the region where current lines
  bend around the defect corner, but they are not $d$-lines as current
  lines connect continuously across them.  Here, $j_{c2} = 0.5\,j_{c1}$
  and $\alpha = \pi/6$.}
  \label{f:jc-larger-outside}
\end{figure}

To proceed with the analysis of the current flow around inhomogeneity
defects, consider the simplified model of
Fig.~\ref{f:defect-inhomogeneous}, where the defect has the shape of a
square located along the edge of the sample. The critical current
density is given by $j_{c2}$ inside the defect and by $j_{c1}$ in the
rest of the sample. The discontinuity lines generated during the
penetration of the magnetic flux are shown in Fig.~\ref{f:jc-larger-outside}.
They are articulated on two isoceles
triangles adjacent to the defect and placed symmetrically about $x =
0$. (One such triangle, composed of the right edge of the defect and the two segments $L_1$ and $L_2$, is represented in Fig.~\ref{f:jc-larger-outside}.) The current lines change direction abruptly along the long side
of each triangle and then either cross the defect vertical border or
go around the defect following a circular path. Each triangle shares
one of its edges with the defect, while its two other edges are given
by
\begin{eqnarray}
  L_1 & : & \quad y = (|x| - b/2)/\tan(\alpha), \\
  L_2 & : & \quad y = (|x| - b/2)/\tan(2 \alpha) + R,
  \label{eq:wedge-lines}
\end{eqnarray}
where $\alpha$ is determined from the ratio of the
sheet current densities,
\begin{eqnarray}
  \label{eq:alpha-Jc1-Jc2}
  \cos 2 \alpha = j_{c2}/j_{c1},
\end{eqnarray}
with $0 \leq \alpha \leq \pi/4$.
The lines $L_1$ and $L_2$ meet at the tip $P_0 = (|x|,y) = (b/2 + R
\sin 2\alpha, R + R \cos 2 \alpha).$ Past $P_0$, the
$d$-line becomes a parabola,
\begin{eqnarray}
  \label{eq:Jc2-Jc1-parabola}
  P & : & \quad y = \frac{(|x|-b/2)^2}{2 R (1-\cos 2\alpha)} + 
\frac{R (1 + \cos 2 \alpha)}{2},
\end{eqnarray}
with a curvature radius
\begin{eqnarray}
  \label{eq:inh-par-a}
  a = R (1 - \cos 2 \alpha).
\end{eqnarray}
By contrast to the lithographically defined indentations studied
in~\cite{Brisbois2016}, the curvature radius $a$ now depends on two
parameters: the depth of the defect, $R$, and the ratio of current
sheet density, $\cos 2 \alpha = j_{c2}/j_{c1}$. More specifically, when the
degree of inhomogeneity is smaller, $\alpha$ is reduced and 
the corresponding parabola branches get closer to the vertical axis.

Another interesting parameter to investigate is the contrast to be
expected when the $d$-line are revealed in a magneto-optical
image. From Eq.~(\ref{eq:contrast}), we have $K_1 = 2 \sin\alpha$
along line $L_1$, $K_2 = 0$ along line $L_2$ (the current lines are
orthogonal to $L_2$), and
$K_3 = 2/(1+(|x|-b/2)^2/(R^2 (1 - \cos 2 \alpha)^2))^{1/2}$ along the
parabola. As a result, the parabolic part is expected to have a higher
contrast near the defect and to be fainter away from the
defect. Moreover, higher contrasts should be observed for 
$\alpha$ closer to $\pi/4$, i.e. for sharp variations of the current densities with
$j_{c2} \ll j_{c1}$. Conversely, defects inducing small discontinuities
in the sheet current densities produce fainter $d$-lines. Thus,
observing both the network of $d$-lines and the contrast should shed
ligth on the geometry and the nature of defects.

\subsection{Power law models}

More refined approaches of the current flow around defects take into
account the actual $E(j)$ constitutive law and lead to a picture that is reminiscent of the ``brick wall'' picture of Fig.~\ref{Fig2}, allowing now some amount of softness of the bricks. Gurevich and Mc Donald\cite{Gurevich1998} introduced an
analytical method based on the hodograph transformation which enabled
the two-dimensional current flow past a planar defect to be evaluated in
superconductors with the isotropic power law $E(j) \propto j^n$. The hodograph method is applied to a superconductor in
steady state with

\begin{eqnarray}
  \label{eq:hodograph-steaty-state}
  \nabla \times \bf{E} = \bf{0}, \quad \nabla\times \bf{H} = 
  {\bf j}({\bf E}),
\end{eqnarray}

\noindent where the electric field derives from an electric potential $\phi$, i.e. $\bf{E}=-\nabla \phi$. With the power law $E \propto j^n$, the resulting equation for $\phi$ is non-linear, but by changing the variables from the Cartesian coordinates $x$ and $y$ to the modulus $E$ and the angle $\theta$ of the electric field, the equation for $\phi$ is linearized. The solution of the problem then consists in solving for $\phi$ in the hodograph space $(E,\theta)$ and performing the inverse mapping of the solution to the real space $(x,y)$. The solution for $\phi$ can be expressed as a series expansion and the inverse mapping to real space can be performed analytically for many geometries~\cite{Gurevich1992,Gurevich1998,Gurevich2000,Friesen2000}. As
shown in~\cite{Gurevich2000}, a linear border defect of depth $a$ forces the current flow to be redistributed over distances
$L_\perp \sim n a$ and $L_\Vert \sim \sqrt{n} a$, respectively measured along a direction  perpendicular or a parallel 
to the current flow. These characteristic lengths satisfy $L_\perp \sim L_{\Vert}^2/ a$, a relationship that is reminiscent
of the parabolic shape of the $d$-line which is predicted in the critical state model, far from the linear defect. 

The hodograph method, with its power law $E(j) \sim j^n$, yields a current flow that shares many similarities with that predicted by the CSM. The current flow is organized in large domains with a prescribed direction of the current density, separated by transition regions  where the current density undergoes sharp local variations. These transitions regions extend in space along curves resembling the CSM $d$-lines. Some of the details of the current flow are different from the CSM results. The authors of~\cite{Gurevich2000}
explain the differences as resulting from the fact that the CSM does not
guarantee that $\nabla \times \bf{E}$ vanishes everywhere in
the sample, and in particular on the $d$-lines, whereas the $j^n$ model does. An interesting illustration of these differences can be observed for the flow of current past a planar defect, evaluated in the limit   $n\to \infty$~\cite{Gurevich2000}: the transition region degenerates into a single parabolic line, $y \simeq x^2/1.94\,a$, with a
radius of curvature $0.96\,a$, i.e.  slightly smaller than the radius $a$ predicted by the CSM
theory.
\paragraph{Surface defects}

As an example of the power law model predictions, the penetration of
magnetic flux around a large circular indentation of about $100~\mu$m
diameter was investigated numerically in Ref.~\cite{Schuster1996}. Note that this case cannot be solved analytically with the hodograph method, as it involves non-linear boundary conditions~\cite{Gurevich2000}. The numerical model
predicts a parabolic $d$-line stemming from the corner
of the indentation, in good accordance with magneto-optical images. It
was also shown that the electric field generated during the magnetic
flux penetration rises very sharply near the indentation, as a result of
the high transit of magnetic flux lines into the specimen. The case of
a semicircular indentation has also been investigated
in Ref.~\cite{Vestgaarden2007}, with a detailed study of the excess
penetration, $\Delta$, defined as the difference between the distance of penetration of the magnetic flux entering from the circular indentation and that for the magnetic flux entering from a straight border (see Fig.~\ref{f:hole-Campbell}).  In the Bean model, suitable for a longitudinal geometry,
the excess penetration $\Delta$ is predicted to be equal to the
indentation radius, $a$. However, in the $E(j) \propto j^n$ model of Ref.~\cite{Vestgaarden2007}, for a thin film,
the excess penetration $\Delta$ is instead observed to be larger than $a$. The excess penetration
$\Delta$ was shown to increase with the radius $a$ and to be a
non-monotonous function of the applied field, with a maximum
increasing with the creep exponent $n$. Its behavior was explained as
resulting from the thin-film geometry, whereby Meissner current
accumulating in front of the indentation lead to a deeper penetration
as their density reaches $j_c$. As initially observed
in Ref.~\cite{Schuster1996}, the electric field is found to be strongly
enhanced near the indentation.

\paragraph{Hole defects}
A similarly interesting case is found in superconducting films with
holes. In these cases, the shielding
currents must also flow around the defect, either towards the center
of the film, thereby producing $d$-lines similar to those observed for
border defects, or they must flow in the space between the
non-conducting hole and the border, where they produce a specific
pattern underneath the hole. This pattern was
interpreted at first as resulting from a second set of discontinuity
lines~\cite{Eisenmenger2001}, but closer scrutiny with the power law
model showed that the pattern results from a local enhancement of the
current density~\cite{Vestgaarden2008}. This pattern can also emerge from
defects with other shapes such as slits or rectangular holes.

Last, in addition to the perturbation in the
current path due to surface defects or holes, the diffusion of
magnetic flux exhibits random fluctuations as a result of disorder and
the dynamics of vortices. The
roughening of magnetic flux fronts penetrating into strongly pinning
YBCO superconducting films with different degrees of edge and bulk
disorder was investigated in Ref.~\cite{Vanderbeek}.

\section{Effect of border defects on thermomagnetic instabilities} 
\label{Sec4}

\subsection{Thermomagnetic instability of superconducting films without defects}

In type-II superconductors, the gradual penetration of magnetic flux in the critical state can be interrupted by thermomagnetic instabilities, which give rise to a drastic degradation of the transport, magnetic, and mechanical properties. This catastrophic breakdown of the performance of materials threatens the safe and stable operation of superconducting devices. During these dramatic events, a large amount of magnetic flux rushes into the superconducting film, accompanied by a sharp local increase in temperature, electric field, and mechanical stress. This event not only deteriorates the electromagnetic performances of superconductors~\cite{Zhao2002,Olsen2007,Chabanenko2022,Choi2007}, but also induces deformation of superconductors~\cite{Jing2015}, potentially even resulting in mechanical damage to superconducting materials~\cite{Baziljevich2014,Shinden2022,Mochizuki2016}.

\begin{figure}
\centering
\includegraphics*[width=0.7\linewidth,angle=0]{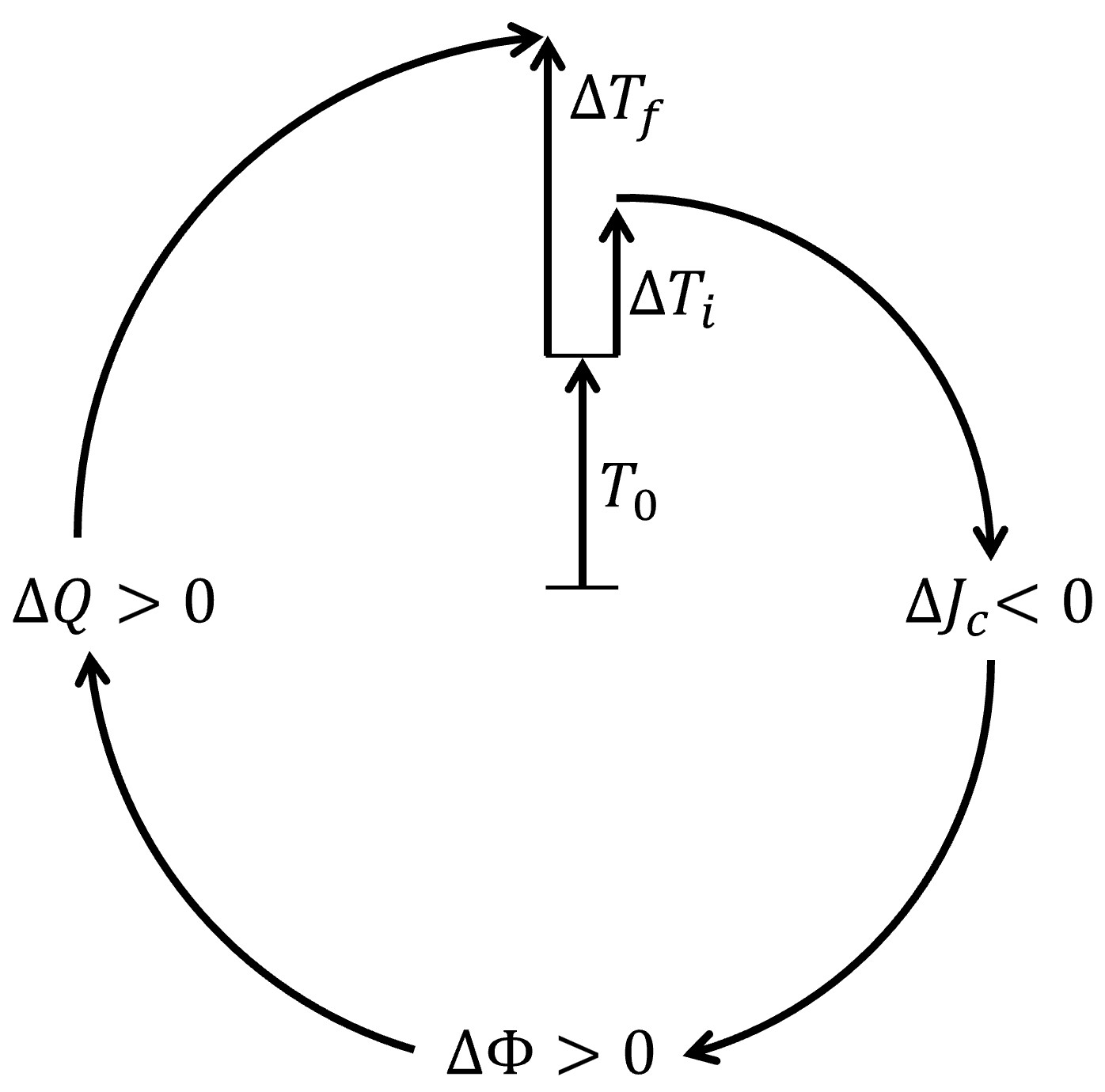}
\caption{Triggering mechanism of a thermomagnetic instability \cite{Chabanenko2000}. $T_0$ represents the global environmental temperature. $\Delta T_i$ is the local rise in temperature caused by sudden magnetic flux motion. As a consequence of this local temperature rise, a decrease of the critical current density $\Delta J_c$ occurs, which in turn leads to further flux displacement $\Delta \Phi$ and heat release $\Delta Q$. Depending on the heat evacuation properties of the sample, the final increase of temperature $\Delta T_f$ can exceed $\Delta T_i$, giving rise to a positive-gain feedback loop triggering a magnetic flux avalanche.}
\label{fig:triggering mechanism}
\end{figure}

The physical mechanism responsible for the thermomagnetic instability can be explained as follows: an initial disturbance of temperature $\Delta T_{i}>0$ locally reduces the critical current density $\Delta J_{c}<0$ due to weakening of the flux pinning, leading to movement and redistribution of magnetic flux within the superconductor. During this process, Joule heat $\Delta Q$ is generated by the vortex displacements, causing a further increase in temperature $\Delta T_{f}$, resulting in a chain reaction ($\Delta T_{i}\uparrow \Rightarrow \Delta J_{c}\downarrow \Rightarrow \Delta Q\uparrow\Rightarrow\Delta T_{f} \uparrow$). The consequence of this process is a positive-gain feedback loop as shown in Fig.~\ref{fig:triggering mechanism}, where an initial minor disturbance develops into a macroscopic thermomagnetic breakdown when the dissipated heat cannot be efficiently evacuated ($\Delta T_{i}<\Delta T_{f}$)~\cite{Wipf1967,Mints1981,Chabanenko2000,Altshuler2004}.

In bulk superconductors, this thermomagnetic instability commonly leads to global flux jumps, sometimes causing the entire superconductor to be heated to the normal state with the displacement of large amounts of flux\cite{Mints1981,Chen1990,Wipf1991,Rakhmanov2004,Kimishima2007,Cun-NatComm}. In superconducting films subjected to an increasing transverse magnetic field, the thermomagnetic instability results in flux avalanches forming complex dendritic patterns. This event takes place in a very short time scale of the order of few nanoseconds, leaving behind a frozen magnetic flux pattern, which can be observed by magneto-optical imaging  (MOI)\cite{Duran1995,Welling2004,Alvarez,Albrecht,Johansen2001,Rudnev2005,Colauto2015,Leiderer1993,Bolz2000,Rudnev2003,Pinheiro2019,Dolan1974,Menghini2005}. An in-depth discussion of these phenomena in low-$T_c$ superconducting films with artificial periodic pinning can be found in \cite{book-oxford,Colauto-review}. Experiments also reveal that the triggering condition and the morphology of the flux avalanche depend on the material and the dimensions of the sample, the working temperature, the magnetic field, and the transport current. A thermomagnetic instability is often triggered during a magnetic-field sweep at an applied field exceeding a threshold magnetic field $H_{th}$ and as long as the temperature is kept below a threshold value $T_{th}$. A broad classification of the avalanche morphology has been pointed out, this being either finger-like or forming branching patterns at extremely high speed (exceeding 100 km/s). This morphology depends on the temperature, with finger-like patterns appearing at low temperatures and branching patterns developing at higher temperatures. Since the nucleation point is not fully reproducible when repeating the experiment under identical conditions, it is difficult to accurately predict the shape and extent of flux avalanches in superconducting films. Various triggering methods, including the use of microwave\cite{Ghigo2007,Awad2011,Lara2017} and focused laser pulses\cite{Leiderer1993,Zhou2020} have been proposed to experimentally stimulate flux avalanches. Recently, Zhou et al. established a hypervelocity MOI system based on the laser-trigger method and multiexposure technology, permitting the effective control of the position where the flux avalanche nucleates\cite{Zhou2020}. In addition, high-speed imaging (30000 frames per second) and high ramping rates (3000 T/s) have been used to obtain real-time imaging of magnetic flux penetration in superconductors\cite{Baziljevich-2012}. Another more straightforward approach (although invasive) for capturing the nanoseconds-scale development of magnetic flux avalanches consists of performing voltage measurements in a superconductor partially coated with a metallic layer, as described in Ref.\cite{Mikheenko2013}.  

To delve deeper into the understanding of triggering mechanisms of thermomagnetic instabilities, analytical modeling based on linear stability analysis of the nucleation stage has been employed to study the threshold for the onset of the flux avalanche. Closed-form expressions for the threshold magnetic field $H_{th}$, temperature $T_{th}$, and electric field $E_{th}$ were derived as functions of material parameters, working temperature, and field ramping rate across various regimes of thermomagnetic instability, and agree well with the experimental observations~\cite{Denisov2006,Denisov2006prb}. Since it remains a challenge to experimentally record the complete and detailed process of thermomagnetic instability from its nucleation stage to the fully developed dendritic pattern, a numerical procedure based on the Fast Fourier Transform (FFT) has been developed to predict the triggering and evolution process of flux avalanches in superconducting films \cite{Vestgarden2018}. 

Several attempts have also been made to control the instabilities, such as coating the superconducting film with a metallic layer\cite{Baziljevich2002,Colauto2010,Stahl2013,Brisbois2017}, inserting artificial defects into the sample\cite{Gheorghe2006,Colauto2013,Motta2014}, adding magnetic stripes\cite{Vlasko-Vlasov2017}, and applying in-plane magnetic fields\cite{Vlasko-Vlasov2016,Colauto2017,Carmo2018}. Vestg{\aa}rden et al. solved the coupled electrodynamics and heat flow in a superconductor-normal metal bilayer to explain the magnetic braking mechanism, using the linearization and numerical simulation of the avalanche dynamics \cite{Vestgarden2013}.

Magnetic flux avalanches have been shown to adversely affect the performance of superconducting resonators\cite{Nulens-2023,Ghigo2007}, giving rise to a noisy response. In addition, it has been suggested that GHz excitation could, in turn, facilitate the triggering of magnetic flux avalanches \cite{Ghigo2007,Awad2011,Ghigo2009,Cuadra}. To some extent, this is not surprising since for typical amplitudes of the RF magnetic field $h_{RF} \sim 0.2$ mT and frequencies $f \sim 5$ GHz, a huge ramp rate $\sim$10$^5$ T/s is obtained. However, direct evidence of RF-enhanced avalanche activity is still lacking. The difficulty lies in the fact that imaging techniques such as MOI are invasive, leading to severe eddy currents in the mirror of the indicator film or modifying the response of the resonant circuit due to different magnetic environmental conditions. Interestingly, early reports suggest that RF (up to 0.3 GHz) superheating is less sensitive to surface defects than DC superheating for both type-I and type-II superconductors \cite{Yogi}.

\subsection{Thermomagnetic instabilities in the presence of border defects} 

As discussed in the previous sections, since border defects strongly modify the flow of the current streamlines over a range much larger than the defect size and significantly affect the electric field generated during the flux penetration \cite{Campbell1972,Brandt1995_1,Brandt-review,Jooss2002}, their influence becomes crucial and in many cases completely dominates the global electromagnetic response. It is well known that a thermomagnetic instability in a superconducting film can occur when the electric field $E$ or the magnetic field $B$ exceed certain threshold value. Consequently, it comes as no surprise that the presence of edge defects may dramatically affect the onset of thermomagnetic instability in the superconducting film. Early studies pointed out that a circular defect should lead to an enhancement of the electric field at the apex of the generated $d$-line parabola and could thus be a nucleation point for thermal instabilities\cite{Schuster1994}. Similar results were predicted for planar defects such as microcracks, low-angle grain boundaries, or defects in the buffer layer\cite{Gurevich2001}.

Mints and Brandt derived a criterion for the onset of a thermomagnetic instability both for a straight edge and an edge with a small indentation by using the linear stability analysis of small coordinate-dependent perturbations \cite{Mints1996}. The stability criterion determining the flux jump field $\mu_0H_{th}$ at the straight edges is given by,

\begin{equation} \label{E1}
\frac{H_{th}\dot{B}_{a}n}{\kappa s^{2}j_{c}}\bigg|\frac{\partial j_{c}}{\partial T}\bigg|=1,
\end{equation}

\noindent where $s$ is determined by the equation $\tan(sd)=h/\kappa s$. $\kappa$ and $h$ are heat conductivity and heat transfer coefficient, respectively, $n$ is the creep exponent, and $d$ is the thickness of the superconducting film. By solving this equation, the dependence of $\mu_0H_{th}$ on the applied magnetic field ramp $\dot B_{a}$ and the critical current density $j_{c}$ can be obtained. Moreover, the stability criterion for flux jumping at a small indentation of radius $a$ is found as

\begin{equation} \label{E2}
\frac{H_{th}^2\dot{B}_{a}n}{2\kappa s^{2}aj_{c}^2}\bigg|\frac{\partial j_{c}}{\partial T}\bigg|=1.
\end{equation}

Comparing the criteria (\ref{E1}) and (\ref{E2}) for the onset of thermomagnetic instability at the straight edge of a film with and without indentation, Mints and Brandt demonstrated that the threshold field or threshold ramp rate for the triggering of the thermomagnetic instability should decrease when introducing a small edge indentation. After that, subsequent studies based on the linear perturbation analysis predicted that edge defects could trigger local thermomagnetic instability due to the amplified local electric field \cite{Schuster1996} and the excess dissipation at defects \cite{Gurevich2000,Gurevich2001}. Moreover, by numerically solving the Maxwell equations with highly nonlinear $E-j$ constitutive relation, the effect of an edge defect on the flux penetration and local electric field in a superconducting film was quantitatively investigated in Ref. \cite{Vestgaarden2007}. The expected enhancement of $E$ due to indentation was also reported in that work, emphasizing the possibility that the edge indentation promotes the onset of flux avalanches in the superconducting film.

\begin{figure}
\centering
\includegraphics*[width=\linewidth,angle=0]{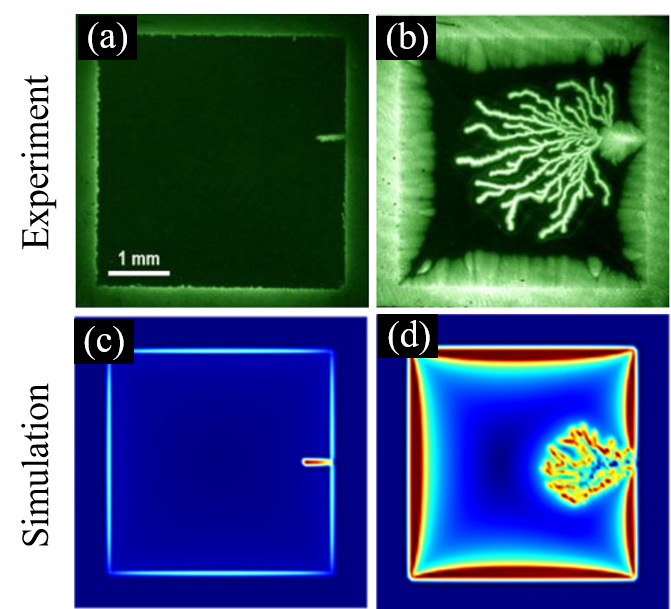}
\caption{MOI experiments (upper row) and numerical simulations (bottom row) for flux patterns in superconducting films with an edge crack. Note that avalanches are preferentially triggered at the crack. Adapted from Refs. \cite{Jing2015,Baziljevich2014}}
\label{fig:instabiltiy_defect}
\end{figure}

Using magneto-optical imaging technique, the predicted phenomenon was observed in YBa$_{2}$Cu$_{3}$O$_{7-\delta}$ by Baziljevich \cite{Baziljevich2014}. The upper panels of Fig.~\ref{fig:instabiltiy_defect} show magneto-optical images of flux distributions in a YBa$_{2}$Cu$_{3}$O$_{7-\delta}$ superconducting film under applied magnetic fields $B_{a}=5$ mT, and $60$ mT, ramping from an initial zero-field at $T_{0}=7$ K. As shown in Fig.~\ref{fig:instabiltiy_defect}(a), a $0.5$ mm long and $80$ $\mu$m wide crack perpendicular to the right edge near the midpoint, facilitates the flux entrance. When the magnetic field is ramped to $60$ mT at a huge rate of $3000$ T/s, a significant dendritic flux avalanche occurs at the crack tip. Such an event did not happen in the defect-free superconducting film, verifying the prediction that edge cracks are prone to triggering thermomagnetic instability. This phenomenon was also confirmed by numerical simulations of MgB$_2$ superconducting films in \cite{Jing2015,Baziljevich2014}, using the Fast Fourier Transform-based iteration scheme (see the lower panels in Fig.~\ref{fig:instabiltiy_defect}). Since then, it has been widely accepted that thermal flux avalanches preferably nucleate at the sites of edge defects.

\begin{figure}
\centering
\includegraphics*[width=\linewidth,angle=0]{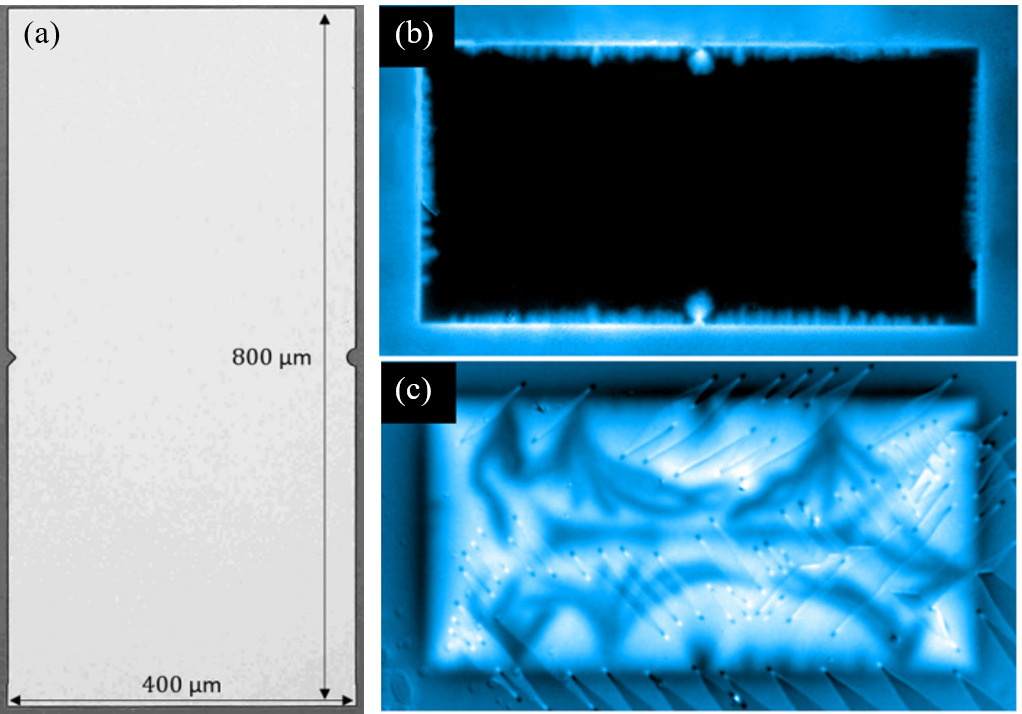}
\caption{(a) Optical microscopy image of the Nb film. MO images of the Nb film at 3.6 K show the flux penetration (b) when a 2 mT magnetic field is applied after ZFC and (c) in a 1 mT field after FC in 12.5 mT. Note that avalanches are not preferentially triggered at indentations, but tend to appear at the smooth borders. Adapted from Ref. \cite{Brisbois2016}.}
\label{fig:instability_smooth edge}
\end{figure}

Interestingly, contrary to the prediction repeated in the numerous studies listed above, edge indentations were not always observed as the preferred nucleation spot for flux avalanches, as reported for Nb superconducting films in Ref. \cite{Brisbois2016}. Fig.~\ref{fig:instability_smooth edge}(a) shows a rectangular sample of $400~\mu\rm m\times800~\mu\rm m$ size with two indentations at the midpoints of the longest sides. When a magnetic field ($2$ mT) is applied after zero-field-cooling down to $3.6$ K, the flux smoothly penetrates the sample without any evidence of flux avalanches, resulting in the expected excess penetration  $\Delta$ and parabolic $d$-lines at the edge indentations, as discussed earlier (see Fig.~\ref{fig:instability_smooth edge}(b)). To corroborate the early predictions, flux avalanches were induced by applying a magnetic field of $1$ mT (ramp rate below 20 mT/s) after field-cooling in $12.5$ mT, as shown in Fig.~\ref{fig:instability_smooth edge}(c). However, these avalanches did not nucleate as expected at the edge indentations; instead, they avoided the edge defects. Moreover, repeated experimental results indicated that this phenomenon was not a random event \cite{Brisbois2016} or dependent on the particular magnetic history, raising the question as to whether edge indentations systematically promote thermomagnetic instabilities.

\begin{figure}
\centering
\includegraphics*[width=\linewidth,angle=0]{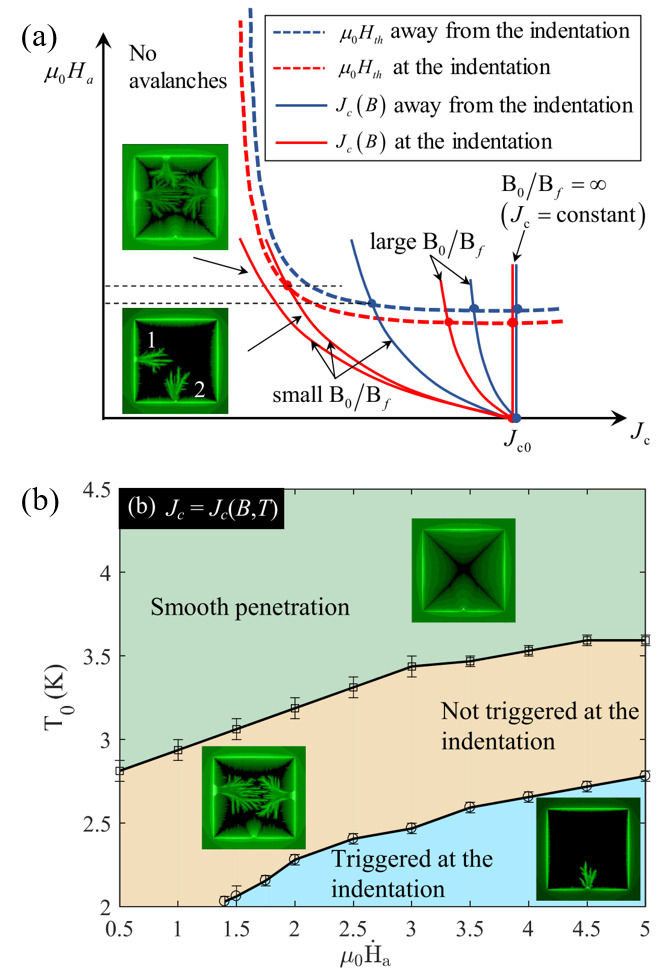}
\caption{(a) Generic curves giving the threshold fields $\mu_{0}H_{th}$ away from the indentation and at the indentation as a function of $J_{c}$. Several $J_{c}(B)$ laws with different $B_{0}/B_{f}$ are shown. (b) Thermomagnetic instability diagram in the $\mu_{0}H_{a}$-$T_{0}$ planes with magnetic-field dependent $J_{c}=J_{c}(B,T)$ with $B_{0}/B_{f}=1$ for a superconducting film with a triangular edge indentation. Adapted from Ref. \cite{Jiang2020}.}
\label{fig:instability_Jiang}
\end{figure}

It is worth noting that the theoretical and numerical calculations mentioned above are based on the critical state model, assuming a constant critical current $j_{c}$ independent of the magnetic field. To address the question of whether edge indentations promote flux avalanches, Jiang et al. \cite{Jiang2020} performed numerical simulations of the thermomagnetic instabilities in a Nb superconducting film by considering a more realistic field-dependent critical current density model.

\begin{equation} \label{E3}
J_{c}=J_{c0}(1-T/T_{c})[B_{0}/(|B|+B_{0})],
\end{equation}

\noindent where $J_{c0}$ is the zero-field critical sheet current and $B_{0}$ reflects the degree of field dependence. By comparing numerical results obtained from field-dependent and field-independent critical currents, they demonstrated that the impact of edge defects on thermomagnetic instabilities depends on the field-dependence of the critical current density. This can be explained by comparing generic curves giving the threshold fields $\mu_{0}H_{th}(J_{c})$ for triggering the first flux avalanche away from the indentation (blue dashed line in Fig.~\ref{fig:instability_Jiang}(a)) or at the indentation (red dashed line in  Fig.~\ref{fig:instability_Jiang}(a)). The red and blue solid lines represent the $J_{c}(\mu_{0}H_{a})$ model for different $B_{0}/B_{f}$, where $B_{f}=\mu_{0}j_{c0}d/\pi$. For a superconducting film with a large $B_{0}/B_{f}$ (e.g., YBa$_{2}$Cu$_{3}$O$_{7-\delta}$, MgB$_2$), one can find that the threshold field is always first reached at the indentation, indicated by the red spot where the solid and dashed red lines intersect. This suggests that flux avalanches in superconducting films of these materials are always triggered first at edge indentations, agreeing well with experimental observations of YBa$_{2}$Cu$_{3}$O$_{7-\delta}$ films \cite{Baziljevich2014} and theoretical predictions based on the constant critical current density model ($B_{0}/B_{f}=\infty$). In contrast to that, in the case of superconducting films with a small value of $B_{0}/B_{f}$ (such as Nb), the fast-rising increase of threshold magnetic field as $J_c$ decreases results in having the first flux avalanche triggered away from the indentation. This situation cannot be explained using a field-independent critical current model. Moreover, the selective triggering of magnetic flux avalanches in Nb film by the edge indentation can be tuned by varying the ramp rate of the applied field $\mu_{0}H_{a}$ and the working temperature $T_{0}$. As shown in Fig.~\ref{fig:instability_Jiang}(b), a phase diagram can be drawn in the $\mu_{0}H_{a}$-$T_{0}$ plane which reveals three distinct regimes of magnetic flux penetration. Two of these regimes are thermomagnetically unstable, with the first flux avalanche being exclusively triggered at the defect location or along a defect-free border. 

\begin{figure}
\centering
\includegraphics*[width=\linewidth,angle=0]{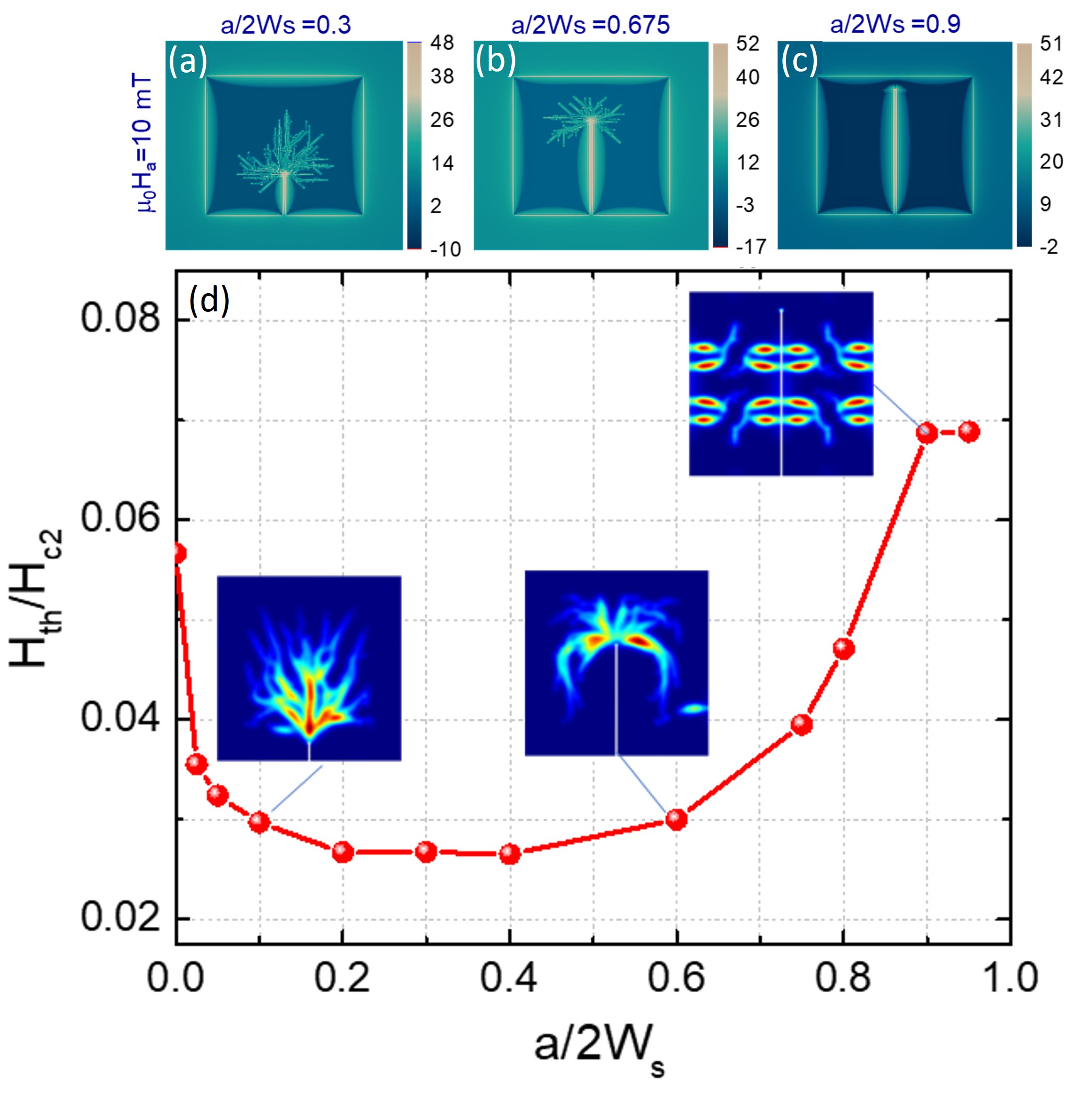}
\caption{(a)-(c) Flux avalanche patterns of the superconducting film with different lengths of an edge crack $a/2W_{s}=0.3$, 0.675, and 0.9. (d) Threshold magnetic field for the first vortex penetration into the film. The inset images illustrate different vortex penetration patterns. Adapted from Ref. \cite{Jing2023}}
\label{fig:instability_crack length}
\end{figure}

More recently, to further emphasize the effects of the edge defect on the flux avalanche in superconducting films, Jing \cite{Jing2023} performed numerical simulations to investigate the threshold condition for the onset of thermomagnetic instabilities, the flux avalanche patterns, the scaling laws and the multifractal spectrum of MgB$_2$ superconducting films with structural defects. It is found that edge cracks sharply decrease the threshold field for the thermomagnetic instability of superconducting films. In addition, low amplitude flux jumps are more frequent than in a sample without cracks. As the crack extends, the threshold field decreases, the jumping frequency gets lower, while the size of the jump gets larger. Remarkably, when the crack extends even further, the threshold field increases to a value exceeding that of the film without a crack. This counterintuitive result is associated with a transition of the thermomagnetic instability triggered at the crack tip to the one triggered at the center of the crack edges. Fig.~\ref{fig:instability_crack length} shows the flux patterns((a)-(c), and the threshold magnetic field for the onset of thermomagnetic instability of the superconducting film (d) with an edge crack of different lengths $a$ with respect to the total width of the sample $2W_s$. The TDGL simulations of the dissipative vortex motion also capture this physical behavior, as depicted in the insets in Fig.~\ref{fig:instability_crack length}(d). These simulations reveal three distinct modes of vortex motion corresponding to varying crack lengths, which effectively reproduce the macroscopic flux dynamics observed during thermomagnetic instability events in the superconducting film.

\section{Conclusion and remaining open questions}
\label{Sec5}

As stated in Section \ref{Section-1}, border defects in superconducting materials act as an entry port for magnetic flux due in part to the current crowding of screening currents and the lowering of edge barriers. In this review, we recalled the main types of surface barriers that can arise in a sample subjected to a magnetic field or a transport current. We discussed the onset of metastability of the Meissner state and the superheating field $H_p$ associated with the Bean-Livingston barrier, as a function of the Ginsburg-Landau parameter $\kappa$ and the geometry of the sample. Bean-Livingston barriers are predominant for thin films ($d < \lambda$) and are sensitive to surface roughness. We also reviewed the more robust geometrical barrier, which is predominant for thicker samples and has a strength that depends on the shape of the sample. Surface barriers were also discussed for anisotropic superconductors and multicomponent superconductors.

Border defects can act as nucleation points for the instability of the Meissner state and thus depress the superheating field $H_p$. We reviewed the variations of $H_p$ for defects of different sizes and the corresponding suppression of the penetration field as a function of the shape of the defect, both within the London and Ginsburg-Landau theories. We further discussed the asymmetry arising between injection and removal of the magnetic flux, we pointed out its use for a rectification effect, and delineated some strategies for increasing $H_p$.
In macroscopic samples in the fully penetrated magnetic state, the impact of border defects is particularly important when the creep exponent is large (i.e. at low temperatures) for which sub-µm scale defects can give rise to abrupt discontinuities in the current direction spanning over macroscopic distances. These discontinuity lines are not crossed by superconducting vortices and leave a clear imprint in the magnetic landscape as captured by magneto-optical imaging. This magnetic marker mainly carries information about the geometrical shape of the defect and the creep exponent for flat borders. The above is true for both bulk samples and thin films. 

In thin films, flux avalanches with dendritic branching develop at low temperatures (typically $T<T_c/2$), as a consequence of a low thermal diffusion (promoting local heating) and high magnetic diffusion (leading to the propagation of magnetic flux). In addition, at low temperatures the critical current increases and therefore the heat released by vortex motion is more substantial than at higher temperatures. These two aspects, combined with the fact that the electric field increases near a border defect, have naturally led to the belief that border defects should act as nucleation spots for vortex avalanches. This has indeed been corroborated within the framework of the magnetic field-independent critical current density model. However, the more realistic field-dependent critical current density model shows that the first flux avalanche can occur either at or away from the defect. This finding is in agreement with experimental results. The selective triggering of magnetic flux avalanches is shown to result from the decrease of the local current density and the increase of the threshold magnetic fields for the first flux avalanche being triggered at the defect. Whether the first avalanche will be nucleated at the defect depends on the defect size, the ramp rate of the applied field, and temperature. In particular, at low rates, avalanches tend to avoid border defects and suggest that magnetic flux avalanches should be less prone to appear along rough borders than along perfectly flat borders. 

Although much has been learned about the influence of border or close-to-border defects on flux penetration, systematic experimental investigations remain scarce, particularly at the microscope scale. In this context, it would be fruitful to resort to microscopic visualization techniques such as SHPM \cite{Kramer-2010}, STM \cite{Timmermans} SOT \cite{Anahory}, NV-centers\cite{Thiel}, MOI \cite{Veshchunov} and the newly emerging technique of scanning quantum vortex microscopy \cite{stolyarov}, able to map the vortex pinning, vortex distribution, and vortex dynamics during flux penetration through border defects. Since the field of view is inversely proportional to the magnification, these techniques should be combined with complementary widefield methods in order to identify the nucleation point along the border. Hence, high-resolution widefield (NV centers, MOI, etc.) are to be privileged over scanning techniques. 

Concerning direct imaging of the dynamic process of flux penetration, the time scales involved in the process (ns) represent an experimental challenge to overcome. The vast majority of the experimental reports so far result from static snapshots of the flux avalanches taken long after they are fully developed. Combined time-resolved magneto-optical imaging and electrical flux detection in type I Pb Thin-Film superconductors has been successfully implemented to visualize individual multiquantum flux tubes during their rapid motion across the superconducting Pb film, yielding spatial and temporal resolution of better than 1 µm and 0.1 µs, respectively \cite{Parisi1985}. Pump-probe laser-based magneto-optical imaging shows a promising and powerful way to cope with the extremely high velocities of flux avalanches in thin films. Alternatively, obtaining thermal maps, strain maps, and electric field maps with high temporal resolution would be highly desirable in order to gain further insights into the initial process of avalanche nucleation. There has been a proposition to use the thermochromic properties of bismuth-substituted yttrium iron garnet thin films to simultaneously visualize the magnetic field and temperature distribution with a temperature sensitivity of 0.2 K \cite{Lee-2017}.

A question that remains unanswered concerns the role of edge barriers (Bean-Livingston, geometrical barriers, etc.) in the onset and morphology of magnetic flux avalanches. It is believed that the nucleation point of thermomagnetic instability lies inside the superconductor, away from the border, where vortices are pinned and therefore edge barriers do not affect the threshold field $H_{th}$. Experimental evidence supporting this interpretation is lacking, in part due to the difficulty of controlling or tuning the properties of edge barriers. New opportunities could emerge by implementing local gating via strong electric fields\cite{DeSimoni, controversy}, particularly effective in thin film HTSC\cite{Ahn,Palau}.

Border defects are ubiquitous, and their impact as magnetic flux gates can have detrimental consequences. A current relevant problem concerns the striation of HTS tapes implemented to reduce the AC losses by as much as one order of magnitude \cite{Pekarcikova}. Various approaches have been tested to create narrow filaments, including dry etching \cite{Marchevsky}, laser scribing \cite{Nast,Terzieva,Sumption}, or laser scribing method combined with wet chemical etching \cite{Machi,Suzuki}. Depending on the method used for creating striation, different border qualities may result and thus impact the magnetic flux penetration in the filamentary tapes. In the same vein, rare-earth barium copper oxide (REBCO) tapes which are fabricated with a fixed width need to be slit down to 1-6 mm for the final user. The slitting process causes micro-cracks in the REBCO and also the underlying buffer layers, which become nucleation sites for continued crack formation and magnetic flux propagation \cite{Hartnett}. In YBCO coated conductors, it has been shown that defects on the conductor edge resulting in delaminated Ag lead to dendritic flux avalanches and high local heating, which cause further Ag delamination \cite{Song}. A recent review on the interface failures of REBCO coated conductors can be found in Ref.\cite{Gao-2023}.

Attempting to comprehensively address the impact of different types of border defects on superconducting samples (inclusions, roughness, local suppression of $T_c$, defects,...), covering the ample space of parameters (geometry, excitation frequency, temperature, ...), and discussing all relevant situations in which surface defects play a role, represents a major challenge. For instance, interesting physics involving the mechanism of flux penetration in superconductors can be found in Refs.\cite{Carmo2016,Carmo2018} in which the authors introduce a superconducting flux injector to allow for controlled injections of magnetic flux into a superconducting film from a predefined location along the edge. Another example is the nucleation of vortices caused by a highly inhomogeneous magnetic field, as discussed in Refs. \cite{Aladyshkin-2011,Ataklti-2012,Aladyshkin-2012}.

The present review intends to provide an overview of the knowledge acquired by the scientific community so far, which in turn should be instrumental in guiding us in the continuous quest for solutions and mitigation strategies to improve the performance of superconducting devices, including radio-frequency cavities, single-photon detectors, and superconducting magnets, to name a few.

\section{Acknowledgements}
\label{Sec6}
The authors acknowledge discussions at the early stages of preparation of this manuscript with Sylvain Blanco Alvarez, Loïc Burger, and Anna Palau and are grateful to Alexey Aladyshkin, Denis Vodolazov, and Grigorii Mikitik for the critical reading and valuable comments and remarks, which have been influential in improving the quality of this report. We also thank Ze Jing and Eli Zeldov, for granting us permission to use their figures in our review. The work of A.V.S. and B.V. was partially supported by the Fonds de la Recherche Scientifique - FNRS under the program EraNet-CHISTERA Grant No. R.8003.21.

\section*{References}

\bibliography{References-tidy} 

\end{document}